%% file: arxiv.tex
\documentclass[format=acmsmall, review=false]{acmart} % Aptara syntax

\usepackage[ruled]{algorithm2e} 

\SetAlFnt{\small}
\SetAlCapFnt{\small}
\SetAlCapNameFnt{\small}
\SetAlCapHSkip{0pt}
\IncMargin{-\parindent}

\acmJournal{CSUR}
\acmVolume{51}
\acmNumber{3}
\acmArticle{0}
\acmYear{2018}
\acmMonth{0}
\copyrightyear{2018}

\acmDOI{0000001.0000001}
\setcopyright{acmcopyright}

\received{October 2016}
\received[revised]{December 2017}
\received[accepted]{January 2018}

\input{common}

\usepackage{background}
\SetBgContents{\fcolorbox{red}{white}{\parbox{\dimexpr\linewidth+12\fboxsep-2\fboxrule}{\itshape\small If you are considering citing this survey, we would appreciate if you could use the following BibTeX entry: \href{http://goo.gl/Hf5Fvc}{http://goo.gl/Hf5Fvc}}}}
\SetBgScale{1}
\SetBgColor{black}
\SetBgAngle{0}
\SetBgOpacity{1}
\SetBgPosition{current page.north}
\SetBgVshift{-0.75cm}
\SetBgHshift{0cm}

\begin{document}

% Title portion
\title{A Survey of Symbolic Execution Techniques}

% Author info
\author{Roberto Baldoni}
\email{baldoni@diag.uniroma1.it}
\author{Emilio Coppa}
\email{coppa@diag.uniroma1.it}
\author{Daniele Cono D'Elia}
\email{delia@diag.uniroma1.it}
\author{Camil Demetrescu}
\email{demetres@diag.uniroma1.it}
\author{Irene Finocchi}
\email{finocchi@di.uniroma1.it}
\affiliation{
	\institution{Sapienza University of Rome}
}

\iffalse
\email{
\affil{\href{http://www.cis.uniroma1.it/}{Cyber Intelligence and Information Security Research Center}, Sapienza}
EMILIO COPPA
\affil{\href{http://season-lab.github.io}{SEASON Lab}, Sapienza University of Rome}
DANIELE CONO D'ELIA
\affil{\href{http://season-lab.github.io}{SEASON Lab}, Sapienza University of Rome}
CAMIL DEMETRESCU
\affil{\href{http://season-lab.github.io}{SEASON Lab}, Sapienza University of Rome}
IRENE FINOCCHI
\affil{\href{http://season-lab.github.io}{SEASON Lab}, Sapienza University of Rome}
}
\fi
% NOTE! Affiliations placed here should be for the institution where the
%       BULK of the research was done. If the author has gone to a new
%       institution, before publication, the (above) affiliation should NOT be changed.
%       The authors 'current' address may be given in the "Author's addresses:" block (below).
%       So for example, Mr. Abdelzaher, the bulk of the research was done at UIUC, and he is
%       currently affiliated with NASA.

\begin{abstract}
Many security and software testing applications require checking whether certain properties of a program hold for any possible usage scenario. For instance, a tool for identifying software vulnerabilities may need to rule out the existence of any backdoor to bypass a program's authentication. One approach would be to test the program using different, possibly random inputs. As the backdoor may only be hit for very specific program workloads, automated exploration of the space of possible inputs is of the essence. Symbolic execution provides an elegant solution to the problem, by systematically exploring many possible execution paths at the same time without necessarily requiring concrete inputs. Rather than taking on fully specified input values, the technique abstractly represents them as symbols, resorting to constraint solvers to construct actual instances that would cause property violations. Symbolic execution has been incubated in dozens of tools developed over the last four decades, leading to major practical breakthroughs in a number of prominent software reliability applications. The goal of this survey is to provide an overview of the main ideas, challenges, and solutions developed in the area, distilling them for a broad audience.
\end{abstract}

%\begin{comment}
\begin{CCSXML} % http://dl.acm.org/ccs.cfm
<ccs2012>
<concept>
<concept_id>10011007.10010940.10010992.10010998.10010999</concept_id>
<concept_desc>Software and its engineering~Software verification</concept_desc>
<concept_significance>500</concept_significance>
</concept>
<concept>
<concept_id>10011007.10010940.10010992.10010998.10011001</concept_id>
<concept_desc>Software and its engineering~Dynamic analysis</concept_desc>
<concept_significance>300</concept_significance>
</concept>
<concept>
<concept_id>10011007.10011074.10011099.10011102.10011103</concept_id>
<concept_desc>Software and its engineering~Software testing and debugging</concept_desc>
<concept_significance>300</concept_significance>
</concept>
<concept>
<concept_id>10002978.10003022</concept_id>
<concept_desc>Security and privacy~Software and application security</concept_desc>
<concept_significance>100</concept_significance>
</concept>
</ccs2012>
\end{CCSXML}

\ccsdesc[500]{Software and its engineering~Software verification}
%\ccsdesc[300]{Software and its engineering~Dynamic analysis}
\ccsdesc[300]{Software and its engineering~Software testing and debugging}
\ccsdesc[100]{Security and privacy~Software and application security}
%\end{comment}

% We no longer use \terms command
%\terms{Design, Algorithms, Performance}

\keywords{Symbolic execution, static analysis, concolic execution, software testing}

%\acmformat{Roberto Baldoni, Emilio Coppa, Daniele Cono D'Elia, Camil Demetrescu,
%and Irene Finocchi, 2016. A survey of symbolic execution techniques.}
% At a minimum you need to supply the author names, year and a title.
% IMPORTANT:
% Full first names whenever they are known, surname last, followed by a period.
% In the case of two authors, 'and' is placed between them.
% In the case of three or more authors, the serial comma is used, that is, all author names
% except the last one but including the penultimate author's name are followed by a comma,
% and then 'and' is placed before the final author's name.
% If only first and middle initials are known, then each initial
% is followed by a period and they are separated by a space.
% The remaining information (journal title, volume, article number, date, etc.) is 'auto-generated'.

\authorsaddresses{Author's addresses: R. Baldoni, E. Coppa, D.C. D'Elia, and C. Demetrescu, Department of Computer, Control, and Management Engineering, Sapienza University of Rome; I. Finocchi, Department of Computer Science, Sapienza University of Rome. E-mail addresses: \{baldoni, coppa, delia, demetrescu\}@diag.uniroma1.it, finocchi@di.uniroma1.it.}

\iffalse
\begin{bottomstuff}
%This work is supported in part by a grant of the Italian Presidency of the Council of Ministers and by the CINI National Laboratory of Cyber Security. % (Consorzio Interuniversitario Nazionale Informatica) 
\end{bottomstuff}
\fi

% Page heads
%\markboth{R. Baldoni, E. Coppa, D. C. D'Elia, C. Demetrescu, and I. Finocchi}{A Survey of Symbolic Execution Techniques}
\renewcommand{\shortauthors}{R. Baldoni et al.}

\maketitle

\input{intro}
\input{executors}
\input{memory}

\input{environment}
\input{explosion}
\input{constraints}
\input{hang}
\input{conclusions}

% Bibliography
%\bibliographystyle{abstract} 
\bibliographystyle{ACM-Reference-Format}
\bibliography{symbolic}

% History dates
%\received{--- 2016}{--- XXXX}{---- XXXX}

\end{document}

% End of v2-acmsmall-sample.tex (March 2012) - Gerry Murray, ACM

%% file: common.tex
\usepackage{listings}
\usepackage{comment}
\usepackage{amsmath}
\usepackage{graphicx}
\usepackage{amssymb}
\usepackage{url}
\usepackage{hyperref}
\usepackage{float}
\usepackage{lipsum}
\usepackage{caption}
\usepackage{subcaption}
\usepackage{adjustbox}
\usepackage{framed}
\usepackage{multirow}
\usepackage{framed}
\usepackage{enumitem}
\usepackage{epigraph}
\usepackage{wasysym} % \brokenvert
\usepackage{wrapfig}

\ifdefined\fullver
\newcommand{\iffullver}[2]{#1}
\else
\newcommand{\iffullver}[2]{#2}
\fi

\definecolor{shadecolor}{rgb}{0.92,0.92,0.92}

\hypersetup{
  colorlinks = true, % colours links instead of ugly boxes
  urlcolor = blue, %  colour for external hyperlinks
  linkcolor = black, % colour of internal links
  citecolor = black, % colour of citations
  pdftitle = {A Survey of Symbolic Execution Techniques},
  pdfauthor= {Roberto Baldoni, Emilio Coppa, Daniele Cono D'Elia, Camil Demetrescu, Irene Finocchi}
}

\newcommand{\myparagraph}[1]{\medskip\noindent{\bf\small #1.} }
\newcommand{\myparagraphnoperiod}[1]{\medskip\noindent{\bf\small #1} }

\newcommand{\mynote}[1]{\ignorespaces} % TODO

\newcommand{\stwoe}{\text{S\textsuperscript{2}E}}

\ifdefined\arxivver
\newcommand{\boxedexample}[1]{
\begin{shaded}
\noindent{\bf\small Example.} #1
\end{shaded}
}
\else
\newcommand{\boxedexample}[1]{
\begin{shaded*}
\noindent{\bf\small Example.} #1
\end{shaded*}
}
\fi

%% file: intro.tex
% !TEX root = main.tex

\epigraph{\textit{``Sometimes you can't see how important something is in its moment, even if it seems kind of important. This is probably one of those times.''}}{(Cyber Grand Challenge highlights from DEF CON 24, August 6, 2016)}

%\vspace{-2.5mm}
\section{Introduction}
\label{se:intro}

Symbolic execution is a popular program analysis technique introduced in the mid '70s to test whether certain properties can be violated by a piece of software~\cite{K-ICRS75,SELECT-ICRS75,K-CACM76,H-TSE77}. Aspects of interest could be that no division by zero is ever performed, no {\tt NULL} pointer is ever dereferenced, no backdoor exists that can bypass authentication, etc. While in general there is no automated way to decide some properties (e.g., the target of an indirect jump), heuristics and approximate analyses can prove useful in practice in a variety of settings, including mission-critical and security applications.

%While in general there is no automated way to decide some properties (think, e.g., of the halting problem), decidable approximations often exist (e.g., ``does a program always terminate within a certain amount of time?''). Such approximations can prove useful in practice in a variety of settings, including mission-critical and security applications.

In a concrete execution, a program is run on a specific input and a single control flow path is explored. Hence, in most cases concrete executions can only under-approximate the analysis of the property of interest. In contrast, symbolic execution can simultaneously explore multiple paths that a program could take under different inputs. This paves the road to sound analyses that can yield strong guarantees on the checked property. 
%\mynote{I: a cosa serve ridirlo? Abbiamo gia' fatto esempi di proprieta' che possono essere verificate}Symbolic execution may answer useful questions on concrete programs like: ``does function {\tt foo(x)} always return a positive value for any possible value of {\tt x}?'' 
The key idea is to allow a program to take on {\em symbolic} -- rather than concrete -- input values. Execution is performed by a {\em symbolic execution engine}, which maintains for each explored control flow path: (i) a first-order Boolean {\em formula} that describes the conditions satisfied by the branches taken along that path, and (ii) a {\em symbolic memory store} that maps variables to symbolic expressions or values. Branch execution updates the formula, while assignments update the symbolic store. A {\em model checker}, typically based on a {\em satisfiability modulo theories} (SMT) solver~\cite{BKM14}, is eventually used to verify whether there are any violations of the property along each explored path and if the path itself is realizable, i.e., if its formula can be satisfied by some assignment of concrete values to the program's symbolic arguments.
%HandbookOfSAT2009

%Variables and control flow paths are associated with expressions and constraints in terms of those symbols during a symbolic execution of the program, and constraints are eventually solved via SMT (satisfiability modulo theories) solvers.

Symbolic execution techniques have been brought to the attention of a heterogeneous audience since DARPA announced in 2013 the Cyber Grand Challenge, a two-year competition seeking to create automatic systems for vulnerability detection, exploitation, and patching in near real-time~\cite{ANGR-SSP16}.
%
% other static program
% which were missed by other program analyses and blackbox testing techniques
More remarkably, symbolic execution tools have been running 24/7 in the testing process of many Microsoft applications since 2008, revealing for instance nearly 30\% of all the bugs discovered by file fuzzing during the development of Windows 7, which other program analyses and blackbox testing techniques missed~\cite{SAGE-QUEUE12}.

In this article, we survey the main aspects of symbolic execution and discuss the most prominent techniques employed for instance in software testing and computer security applications. Our discussion is mainly focused on {\em forward} symbolic execution, where a symbolic engine analyzes many paths simultaneously starting its exploration from the main entry point of a program.
%its extensive usage in software testing and computer security applications\mynote{[D] this should change}, where software vulnerabilities can be found by symbolically executing programs at the level of either source or binary code. 
%A different approach is symbolic {\em backward} execution, where exploration is started from a specific point of the program (e.g., an {\tt assert} statement) and the engine proceeds backward, trying to reconstruct a valid path from an entry point of the program. Since forward symbolic execution is the mainline technique in literature, throughout this article we will always refer to this approach when using the term symbolic execution. Nonetheless, some benefits offered by symbolic backward execution will be pointed out when relevant for the discussion.

We start with a simple example that highlights many of the fundamental issues addressed in the remainder of the article.

% --------------------------------------------------------------------------------------------------------------------
\subsection{A Warm-Up Example}
\label{symbolic-execution-example}

\begin{figure}[t]
\begin{center}
\begin{tabular}{c}
\begin{lstlisting}[basicstyle=\ttfamily\scriptsize]
1.  void foobar(int a, int b) {
2.     int x = 1, y = 0;
3.     if (a != 0) {
4.        y = 3+x;
5.        if (b == 0)
6.           x = 2*(a+b);
7.     }
8.     assert(x-y != 0);
9.  }
\end{lstlisting}
\end{tabular}
\end{center}
\vspace{-2mm}
\caption{Warm-up example: which values of \texttt{a} and \texttt{b} make the \texttt{assert} fail?}
\label{fig:example-1}
%\vspace{-1.5mm}
\end{figure}

%\revedit{in the common 4-byte representation}
Consider the C code of Figure~\ref{fig:example-1} and assume that our goal is to determine which inputs make the {\tt assert} at line 8 of function \texttt{foobar} fail. Since each 4-byte input parameter can take as many as $2^{32}$ distinct integer values, the approach of running concretely function \texttt{foobar} on randomly generated inputs will unlikely pick up exactly the assert-failing inputs.
%Techniques such as random testing could generate bottomless input tests for this function. 
%However, it is unlikely that exactly the assert-failing inputs would be randomly picked up\mynote{Fuzzing?}. 
By evaluating the code using symbols for its inputs, instead of concrete values, symbolic execution overcomes this limitation and makes it possible to reason on {\em classes of inputs}, rather than single input values. 

In more detail, every value that cannot be determined by a static analysis of the code, such as an actual parameter of a function or the result of a system call that reads data from a stream, is represented by a symbol $\alpha_i$. At any time, the symbolic execution engine maintains a state $(stmt,~\sigma,~\pi)$ where:

\begin{itemize}[itemsep=2pt]

\item $stmt$ is the next statement to evaluate. For the time being, we assume that $stmt$ can be an assignment, a conditional branch, or a jump (more complex constructs such as function calls and loops will be discussed in  Section~\ref{se:path-explosion}).

%\item $\sigma$ is a {\em symbolic store} that associates program variables with expressions over \mynote{[D] $\alpha_i$ also concrete?} concrete and symbolic values $\alpha_i$.

\item $\sigma$ is a {\em symbolic store} that associates program variables with either expressions over concrete values or symbolic values $\alpha_i$.

\item $\pi$ denotes the {\em path constraints}, i.e., is a formula that expresses a set of assumptions on the symbols $\alpha_i$ due to branches taken in the execution to reach $stmt$. At the beginning of the analysis, $\pi=true$.

\end{itemize}

\noindent Depending on $stmt$, the symbolic engine changes the state as follows:

\begin{itemize}[topsep=3pt,itemsep=2pt] % TODO
  \item The evaluation of an assignment $x=e$ updates the symbolic store $\sigma$ by associating $x$ with a new symbolic expression $e_s$. We denote this association with $x\mapsto e_s$, where $e_s$ is obtained by evaluating $e$ in the context of the current execution state and  can be any expression involving unary or binary operators over symbols and concrete values.
  
%   $\alpha_i = e$: when an expression $e$ is assigned to a symbol $\alpha_i$, $pc$ is extended by adding a constraint on $\alpha_i$:
%    \[ pc \gets pc \wedge \alpha_i = e\]
%  where $e$ can be any expression, involving unary or binary operators, over symbols and constants.

  \item The evaluation of a conditional branch ${\tt if}~e~{\tt then}~s_{true}~{\tt else}~s_{false}$ affects the path constraints $\pi$. The symbolic execution is forked by creating two execution states with path constraints $\pi_{true}$ and $\pi_{false}$, respectively, which correspond to the two branches: $\pi_{true}=\pi \wedge e_s$ and $\pi_{false}=\pi \wedge \neg e_s$, where $e_s$ is a symbolic expression obtained by evaluating $e$. 
%        \[ (s_{true}, pc_{true}) \text{ where } pc_{true} = pc \wedge e \]
%        \[ (s_{false}, pc_{false}) \text{ where } pc_{false} = pc \wedge \neg e \]
    Symbolic execution independently proceeds on both states.

  \item The evaluation of a jump {\tt goto} $s$ updates the execution state by advancing the symbolic execution to statement $s$. 
\end{itemize}

%\subsection{Example}
%\label{symbolic-execution-example}

%\begin{figure}[t]
%  \centering
%  \includegraphics[width=1.0\columnwidth]{images/example} 
%  \caption{Symbolic execution tree of the function {\tt foobar}. Each execution state is labeled with an alphabet letter. Side effects on execution states are highlighted in gray. Leaves are evaluated against division by zero error. For the sake of presentation the conjunction of constraints is shown as a list of constraints. }
%  \label{fig:example-symbolic-execution}
%\end{figure}

\begin{figure}[t]
  \centering
  \includegraphics[width=0.975\columnwidth]{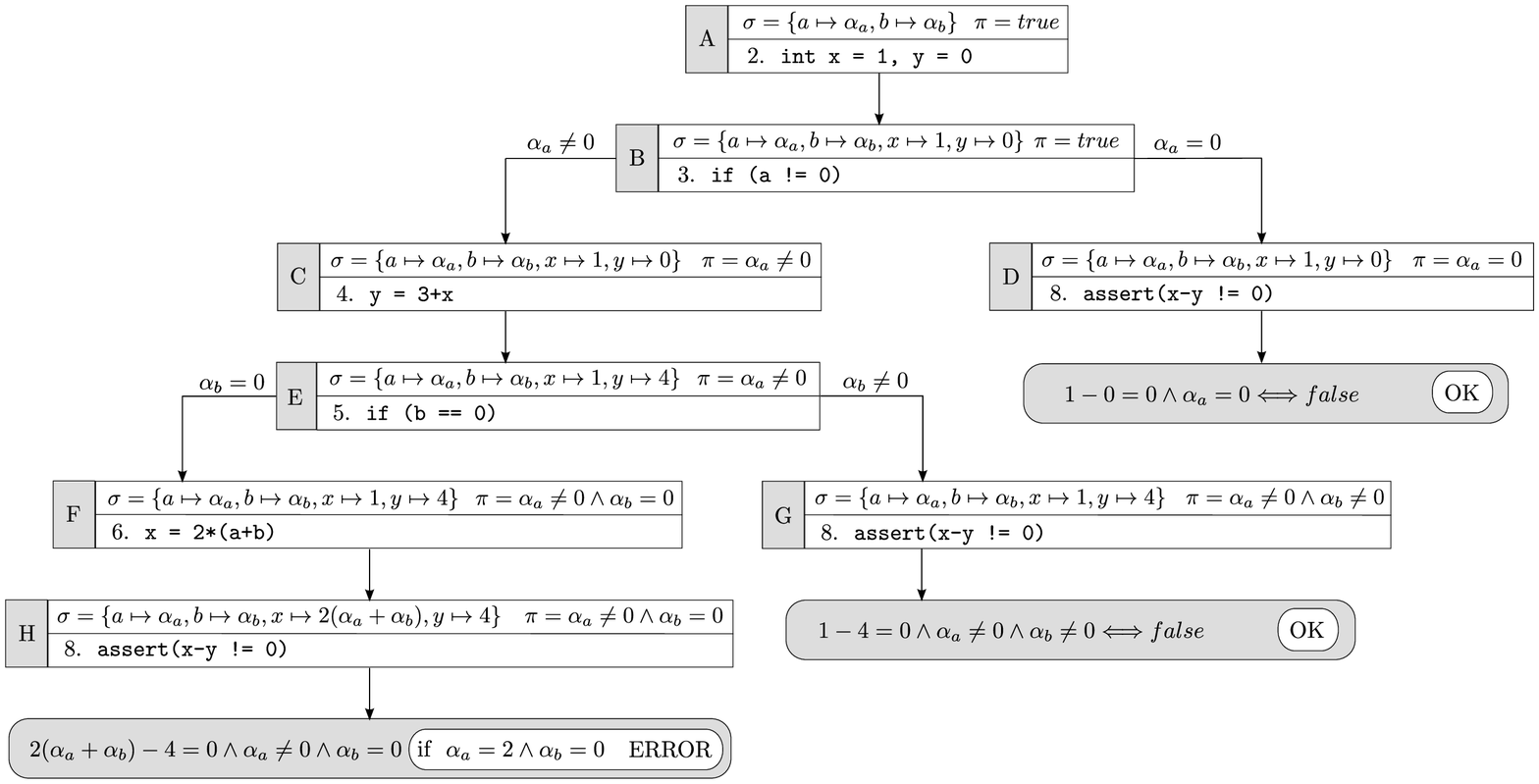} 
  \caption{Symbolic execution tree of function {\tt foobar} given in Figure~\ref{fig:example-1}. Each execution state, labeled with an upper case letter, shows the statement to be executed, the symbolic store $\sigma$, and the path constraints $\pi$. Leaves are evaluated against the condition in the {\tt assert} statement. }
%For the sake of presentation the conjunction of constraints is shown as a list of constraints. }
  \label{fig:example-symbolic-execution}
  %\vspace{-1mm}
\end{figure}

\noindent A symbolic execution of function {\tt foobar}, which can be effectively represented as a tree, is shown in Figure~\ref{fig:example-symbolic-execution}. Initially (execution state $A$) the path constraints are {\tt true} and input arguments {\tt a} and {\tt b} are associated with symbolic values. 
After initializing local variables {\tt x} and {\tt y} at line 2, the symbolic store is updated by associating {\tt x} and {\tt y} with concrete values 1 and 0, respectively (execution state $B$). Line 3 contains a conditional branch and the execution is forked: depending on the branch taken, a different statement is evaluated next and different assumptions are made on symbol $\alpha_a$ (execution states $C$ and $D$, respectively). In the branch where $\alpha_a\neq 0$, variable {\tt y} is assigned with ${\tt x}+3$, obtaining $y\mapsto 4$ in state $E$ because $x\mapsto 1$ in state $C$. In general, arithmetic expression evaluation simply manipulates the symbolic values.
After expanding every execution state until the {\tt assert} at line 8 is reached on all branches, we can check which input values for parameters {\tt a} and {\tt b} can make the {\tt assert} fail. By analyzing execution states $\{D,G,H\}$, we can conclude that only $H$ can make {\tt x-y = 0} true. The path constraints for $H$ at this point implicitly define the set of inputs that are unsafe for {\tt foobar}. 
In particular, any input values such that:
 \[ 2(\alpha_a+\alpha_b)-4 = 0 \wedge \alpha_a \neq 0 \wedge \alpha_b = 0 \]
will make {\tt assert} fail. An instance of unsafe input parameters can be eventually determined by invoking an {\em SMT solver}~\cite{BKM14} to solve the path constraints, which in this example would yield $a = 2$ and $b = 0$. % HandbookOfSAT2009

%Notice\mynote{Say earlier?} that a constraint solver is also needed when evaluating the satisfiability of branch conditions.

% --------------------------------------------------------------------------------------------------------------------
\subsection{Challenges in Symbolic Execution}
\label{example-discussion}

In the example discussed in Section~\ref{symbolic-execution-example} symbolic execution can identify {\em all} the possible unsafe inputs that make the {\tt assert} fail. This is achieved through an exhaustive exploration of the possible execution states. From a theoretical perspective, exhaustive symbolic execution provides a {\em sound} and {\em complete} methodology for any decidable analysis. Soundness prevents false negatives, i.e., all possible unsafe inputs are guaranteed to be found, while completeness prevents false positives, i.e.,  input values deemed unsafe are actually unsafe. As we will discuss later on, exhaustive symbolic execution is unlikely to scale beyond small applications. Hence, in practice we often settle for less ambitious goals, e.g., by trading soundness for performance.

Challenges that symbolic execution has to face when processing real-world code can be significantly more complex than those illustrated in our warm-up example. Several observations and questions naturally arise:

\begin{itemize}[itemsep=1mm]
%%%
\item \noindent {\em Memory}: how does the symbolic engine handle pointers, arrays, or other complex objects? Code manipulating pointers and data structures may give rise not only to symbolic stored data, but also to addresses being described by symbolic expressions.
%Any arbitrarily complex object can be regarded as an array of bytes and each byte associated with a distinct symbol. However, when possible, exploiting structural properties of the data may be more convenient: for instance, relational bounds on the class fields in object-oriented languages could be used for refining the search performed by symbolic execution.
%%%
\item {\em Environment}: how does the engine handle interactions across the software stack? Calls to library and system code can cause side effects, e.g., the creation of a file or a call back to user code, that could later affect the execution and must be accounted for. However, evaluating any possible interaction outcome may be unfeasible.
%: it would give rise to a large number of states, while only a fraction of them can \mynote{[D] likely?}actually happen in a non-symbolic scenario.
%%\mytempedit{Also, third-party closed-source components and popular frameworks (e.g., Java Swing and Android) pose further challenges to an executor, for instance because of the control flows occurring within them through callbacks.}\mynote{CD: may be dropped if we run out of space}
% Real-world applications constantly interact with the environment (e.g., the file system or the network) through libraries and system calls. These interactions may cause side effects (such as the creation of a file) that could later affect the execution and must be therefore taken into account. Evaluating any possible interaction outcome is generally unfeasible: it could generate a large number of execution states, of which only a small number can actually happen in a non-symbolic scenario. %A typical strategy is to consider popular library and system routines and create models that can help the symbolic engine analyze only significant outcomes.
%%%
  \item {\em State space explosion}: how does symbolic execution deal with path explosion?
%\mynote{[D] I felt it was too long and loop-centric} 
Language constructs such as loops might exponentially increase the number of execution states. It is thus unlikely that a symbolic execution engine can exhaustively explore all the possible states within a reasonable amount of time. %In practice, heuristics are used to guide exploration and prioritize certain states first (e.g., to maximize code coverage). In addition, 
%In practice, several heuristics must be exploited to prioritize evaluation of some states, hoping to still be able to spot interesting things. Moreover, the symbolic execution engine should include efficient mechanism for efficiently evaluating in parallel different execution states without running out of computational resources.
%%%
  \item {\em Constraint solving}: what can a constraint solver do in practice?
  %{\em What is a constraint solver in practice}? \\
SMT solvers can scale to complex combinations of constraints over hundreds of variables. However, constructs such as non-linear arithmetic pose a major obstacle to efficiency.
%Constraint solvers suffer from a number of limitations. They can typically handle complex constraints in a reasonable amount of time only if they are made of linear expressions over their constituents.
%Constraint solvers suffer from a number of limitations. They can typically handle complex constraints in a reasonable amount of time only if they are made of linear expressions over their constituents. %Symbolic execution engines normally implement a number of optimizations to make queries as much {\em solver-friendly} as possible, for instance by splitting queries into independent components to be processed separately or by performing algebraic simplifications.
%%%
\iffullver{  \item {\em Binary code}: what issues can arise when symbolically executing binary code?
  %what are the disadvantages of symbolically executing binary code?
 While the warm-up example of Section~\ref{symbolic-execution-example} is written in C, in several scenarios binary code is the only available representation of a program. However, having the source code of an application can make symbolic execution significantly easier, as it can exploit high-level properties (e.g., object shapes) that can be inferred statically by analyzing the source code.
 }{}
%(e.g., the maximum size of a buffer or the number of iterations for a loop).
%%%   
\end{itemize}
%Depending on the specific application context of symbolic execution

\noindent Depending on the specific context in which symbolic execution is used, different choices and assumptions are made to address the questions highlighted above. Although these choices typically affect soundness or completeness, in several scenarios a partial exploration of the space of possible execution states may be sufficient to achieve the goal (e.g., identifying a crashing input for an application) within a limited time budget.

%different choices and assumptions are made to address the above questions. Although soundness and completeness of symbolic execution may be negatively affected by these choices, there are several application scenarios where a partial exploration of the possible execution states is sufficient for reaching the ultimate goal (e.g., identify a single input that crashes an application).

% --------------------------------------------------------------------------------------------------------------------
\subsection{Related Work}
\label{ss:related-surveys}

Symbolic execution has been the focus of a vast body of literature. As of August 2017, Google Scholar reports 742 articles that include the exact phrase ``symbolic execution'' in the title. Prior to this survey, other authors have contributed technical overviews of the field, such as \cite{PV-JSTTT09} and \cite{CS-CACM13}. \cite{CHEN20131758} focuses on the more specific setting of automated test generation: it provides a comprehensive view of the literature, covering in depth a variety of techniques and complementing the technical discussions with a number of running examples.
%Besides  complementing the technical discussions with a number of running examples, it covers in depth recent techniques for key aspects such as memory modelling, environment interaction, path explosion, and constraint solving.

% --------------------------------------------------------------------------------------------------------------------
\subsection{Organization of the Article}
\label{ss:article-organization}

The remainder of this article is organized as follows. In Section~\ref{se:executors} we discuss the overall principles and evaluation strategies of a symbolic execution engine. Section~\ref{memory-model} through Section~\ref{se:constraint-solving} address the key challenges that we listed in Section~\ref{example-discussion}, while Section~\ref{se:hang} discusses how recent advances in other areas could be applied to enhance symbolic execution techniques. Concluding remarks are addressed in Section~\ref{se:conclusions}.

%% file: executors.tex
% !TEX root = main.tex

\section{Symbolic Execution Engines}
\label{se:executors}

In this section we describe some important principles for the design of symbolic executors and crucial tradeoffs that arise in their implementation. Moving from the concepts of concrete and symbolic runs, we also introduce the idea of {\em concolic} execution.

\subsection{Mixing Symbolic and Concrete Execution}
\label{ss:concrete-concolic-symbolic}

\begin{figure}[t]
\centering
%\vspace{-0.75mm}
\includegraphics[width=0.32\columnwidth]{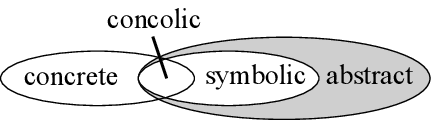}  % TODO 0.34
\caption{Concrete and abstract execution machine models.}
\label{fig:concrete-symbolic}
\vspace{-1.5mm}
\end{figure}

As shown in the warm-up example (Section~\ref{symbolic-execution-example}), a symbolic execution of a program can generate -- in theory -- all possible control flow paths that the program could take during its concrete executions on specific inputs. While modeling all possible runs allows for very interesting analyses, it is typically unfeasible in practice, especially on real-world software. %, for a variety of reasons.

A main limitation of classical symbolic execution is that it cannot explore feasible executions that would result in path constraints that cannot be dealt with~\cite{CS-CACM13}. Loss of soundness originates from external code not traceable by the executor, as well as from complex constraints involving, e.g., non-linear arithmetic or transcendental functions. As the time spent in constraint solving is a major performance barrier for an engine, solvability can be intended in the absolute sense, but as in efficiency too. Also, practical programs are typically not self-contained: implementing a symbolic engine able to statically analyze the whole software stack can be rather challenging given the difficulty in accurately evaluating any possible side effect during execution.
A fundamental idea to cope with these issues and to make symbolic execution feasible in practice is to mix concrete and symbolic execution: this is dubbed {\em concolic execution}, where the term concolic is a portmanteau of the words ``concrete'' and ``symbolic'' (Figure~\ref{fig:concrete-symbolic}). This general principle has been explored along different angles, discussed in the remainder of this section.
%
%issues (\ref{it:library-call-issue}), (\ref{it:third-party-issue}), and (\ref{it:smt-solver-issue})

\newcommand{\dse}{DSE} % TODO move/drop
\myparagraph{Dynamic Symbolic Execution} One popular concolic execution approach, known as {\em dynamic symbolic execution} (\dse) or {\em dynamic test generation}~\cite{DART-PLDI05}, is to have concrete execution drive symbolic execution. This technique can be very effective in mitigating the issues above. In addition to the symbolic store and the path constraints, the execution engine maintains a concrete store $\sigma_c$. After choosing an arbitrary input to begin with, it executes the program both concretely and symbolically by simultaneously updating the two stores and the path constraints. Whenever the concrete execution takes a branch, the symbolic execution is directed toward the same branch and the constraints extracted from the branch condition are added to the current set of path constraints. In short, the symbolic execution is driven by a specific concrete execution. As a consequence, the symbolic engine does not need to invoke the constraint solver to decide whether a branch condition is (un)satisfiable: this is directly tested by the concrete execution. In order to explore different paths, the path conditions given by one or more branches can be negated and the SMT solver invoked to find a satisfying assignment for the new constraints, i.e., to generate a new input. This strategy can be repeated as much as needed to achieve the desired coverage.

\vspace{-2pt} 
\boxedexample{
Consider the C function in Figure~\ref{fig:example-1} and suppose to choose $a = 1$ and $b = 1$ as input parameters. Under these conditions, the concrete execution takes path $A\leadsto B\leadsto C\leadsto E\leadsto G$ in the symbolic tree of Figure~\ref{fig:example-symbolic-execution}. Besides the symbolic stores shown in Figure~\ref{fig:example-symbolic-execution}, the concrete stores maintained in the traversed states are the following:
  \begin{enumerate}
  
  \item[]$-$~~$\sigma_c=\{a\mapsto 1,~b\mapsto 1\}$ in state $A$;
  \item[]$-$~~$\sigma_c=\{a\mapsto 1,~b\mapsto 1,~x\mapsto 1,~y\mapsto 0\}$ in states $B$ and $C$;
  \item[]$-$~~$\sigma_c=\{a\mapsto 1,~b\mapsto 1,~x\mapsto 1,~y\mapsto 4\}$ in states $E$ and $G$.
  
  \end{enumerate}  
%Stores and path constraints maintained by the concolic run are shown in Figure~\ref{fig:example-concolic-execution}. 
After checking that the \texttt{assert} conditions at line 8 succeed, we can generate a new control flow path by negating the last path constraint, i.e., $\alpha_b\neq 0$. The solver at this point would generate a new input that satisfies the constraints $\alpha_a\neq 0\,\wedge\, \alpha_b=0$ (for instance $a = 1$ and $b = 0$) and the execution would continue in a similar way along the path $A\leadsto B\leadsto C\leadsto E\leadsto F$. %
}

\vspace{-2pt}  % TODO was +2mm
% generation search algorithm discussed in -> generational search of
% and, for this reason, -> hence
%\noindent
Although \dse\ uses concrete inputs to drive the symbolic execution toward a specific path, it still needs to pick a branch to negate whenever a new path has to be explored. Notice also that each concrete execution may add new branches that will have to be visited. Since the set of non-taken branches across all performed concrete executions can be very large, adopting effective search heuristics (Section~\ref{ss:heuristics}) can play a crucial role. For instance, {\sc DART}~\cite{DART-PLDI05} chooses the next branch to negate using a depth-first strategy. Additional strategies for picking the next branch to negate have been presented in literature. For instance, the {\em generational search} of {\sc SAGE}~\cite{SAGE-NDSS08} systematically yet partially explores the state space, maximizing the number of new tests generated while also avoiding redundancies in the search. This is achieved by negating constraints following a specific order and by limiting the backtracking of the search algorithm. Since the state space is only partially explored, the initial input plays a crucial role in the effectiveness of the overall approach.  The importance of the first input is similar to what happens in traditional {\em black-box fuzzing}; hence, symbolic engines such as {\sc SAGE} are often referred to as {\em white-box fuzzers}.

The symbolic information maintained during a concrete run can be exploited by the engine to obtain new inputs and explore new paths. The next example shows how DSE \mynote{[D] deals with?} can handle invocations to external code that is not symbolically tracked by the concolic engine. Use of concrete values to aid constraint solving will be discussed in Section~\ref{se:constraint-solving}.

%The symbolic information maintained during a concrete run can be exploited by the execution engine, for instance, to obtain new inputs and explore new control flow paths. The next example shows how dynamic symbolic execution \mynote{[D] deals with?} can handle invocations to external code that is not symbolically tracked by the concolic engine.
%However, our previous example has not motivated in practice the two main benefits given by concolic execution: (a) use of the concolic store to help a SMT solver to efficiently handle non-linear constraints, (b) symbolic execution of a piece of code that contains invocations to external code that is not symbolically tracked by the concolic engine. While the first benefit will be better explained in Section~\ref{se:constraint-solving}, we now use the next example to explore the second benefit, pinpointing also some critical disadvantages that come with it. 

\vspace{-2pt}
\boxedexample{
Consider function {\tt foo} in Figure~\ref{fig:example-concolic-problems}a and suppose that {\tt bar} is not symbolically tracked by the concolic engine (e.g., it could be provided by a third-party component, written in a different language, or analyzed following a black-box approach). Assuming that $x = 1$ and $y = 2$ are randomly chosen as the initial input parameters, the concolic engine executes {\tt bar} (which returns $a = 0$) and skips the branch that would trigger the error statement. At the same time, the symbolic execution tracks the path constraint $\alpha_y \geq 0$ inside function {\tt foo}. Notice that branch conditions in function {\tt bar} are not known to the engine. To explore the alternative path, the engine negates the path constraint of the branch in {\tt foo}, generating inputs, such as $x = 1$ and $y = -4$, that actually drive the concrete execution to the alternative path. With this approach, the engine can explore both paths in {\tt foo} even if {\tt bar} is not symbolically tracked. 

A variant of the previous code is shown in Figure~\ref{fig:example-concolic-problems}b, where function {\tt qux} -- differently from {\tt foo} -- takes a single input parameter but checks the result of {\tt bar} in the branch condition. Although the engine can track the path constraint in the branch condition tested inside {\tt qux}, there is no guarantee that an input able to drive the execution toward the alternative path is generated: the relationship between $a$ and $x$ is not known to the concolic engine, as {\tt bar} is not symbolically tracked. In this case, the engine could re-run the code using a different random input, but in the end it could fail to explore one interesting path in {\tt qux}. 

A related issue is presented by Figure~\ref{fig:example-concolic-problems}c. We observe a {\em path divergence} when inputs generated for a predicted path lead execution to a different path. In general, this can be due to symbol propagation not being tracked, resulting in inaccurate path constraints, or to imprecision in modeling certain (e.g., bitwise, floating-point) operations in the engine. In the example, function {\tt baz} invokes the external function {\tt abs}, which performs a side effect on $x$ by assigning it with its absolute value. Choosing $x = 1$ as the initial concrete value, the concrete execution does not trigger the error statement, but the concolic engine tracks the path constraint $\alpha_x \geq 0$ due to the branch in {\tt baz}, trying to generate a new input by negating it. However the new input, e.g., $x = -1$, does not trigger the error statement due to the (untracked) side effects of {\tt abs}. Interestingly, the engine has no way of detecting that no input can actually trigger the error.
% {\tt abs}, which simply computes the absolute value of a number
%In this case, after generating a new input the engine detects a {\em path divergence}: a concrete execution that does not follow the predicted path. Interestingly, in this example no input could actually trigger the error, but the engine is not able to detect this property.
}

\begin{figure*}[t]
  \vspace{-2.5mm}
  %\centering
  \begin{subfigure}[t]{.33\textwidth}
    \begin{lstlisting}[basicstyle=\ttfamily\scriptsize]
void foo(int x, int y) {
   int a = bar(x);
   if (y < 0) ERROR;
} 
\end{lstlisting}
%int bar(int z) {
%   if (z == 23) return 1;
%   return 0;
%}
    \vspace{-4mm}
    \caption{}
  \end{subfigure}%
  %
  %\hspace{-2mm}
  %
  \begin{subfigure}[t]{.33\textwidth}
    %\vspace{5mm}
    \begin{lstlisting}[basicstyle=\ttfamily\scriptsize]
void qux(int x) {
   int a = bar(x);
   if (a > 0) ERROR;
} 
\end{lstlisting}
%int bar(int z) {
%   if (z == 23) return 1;
%   return 0;
%}
    \vspace{-4mm}
    \caption{}
  \end{subfigure}%
  %
  %\hspace{-2mm}
  %
  \begin{subfigure}[t]{.33\textwidth}
    %\vspace{5mm}
    \begin{lstlisting}[basicstyle=\ttfamily\scriptsize]
void baz(int x) {
   abs(&x);
   if (x < 0) ERROR;
}   
\end{lstlisting}
%void abs(int * z) {
%   if (*z < 0) *z = -(*z);
%}
    \vspace{-4mm}
    \caption{}
  \end{subfigure}%

  \vspace{-2mm}
  \caption{Concolic execution: (a) testing of function {\tt foo} even when {\tt bar} cannot be symbolically tracked by an engine, (b) example of false negative, and (c) example of a path divergence, where \texttt{abs} drops the sign of the integer at \texttt{\&x}.
   %(b) Example of a missed path: assuming that function {\tt bar} is not symbolically tracked, then it is unlikely that the engine will be able to generate an input that trigger execution of function {\tt error} inside function {\tt qux}. (c) Example of a path divergence: assuming that function {\tt abs} is not tracked, then the concolic engine may try to generate fruitlessly an input able to trigger execution of function {\tt error} in function {\tt baz}.
   }
  \label{fig:example-concolic-problems}
  \vspace{-3mm}
\end{figure*}

%\noindent 
As shown by the example, false negatives (i.e., missed paths) and path divergences are notable downsides of dynamic symbolic execution. \dse\ trades soundness for performance and implementation effort: false negatives are possible, because some program executions -- and therefore possible erroneous behaviors -- may be missed, leading to a {\em complete}, but {\em under-approximate} form of program analysis. Path divergences have been frequently observed in literature: for instance,~\cite{SAGE-NDSS08} reports rates over $60\%$. \cite{CLH-SCN15} presents an empirical study of path divergences, analyzing the main patterns that contribute to this phenomenon. External calls, exceptions, type casts, and symbolic pointers are pinpointed as critical aspects of concolic execution that must be carefully handled by an engine to reduce the number of path divergences.

%\mytempedit{
%To mitigate the problem of false negatives, several works have exploited variants of concolic execution or different ways of mixing concrete and symbolic runs. \cite{DBF-ISSTA16} presents a framework for the specification of mixed concrete/symbolic policies, allowing for a better understanding of their impact and trade-offs in terms of correctness, completeness and efficiency of the approach.} 

\myparagraph{Selective Symbolic Execution}
{\sc \stwoe}~\cite{CKC-TOCS12} takes a different approach to mix symbolic and concrete execution based on the observation that one might want to explore only some components of a software stack in full, not caring about others. {\em Selective symbolic execution} carefully interleaves concrete and symbolic execution, while keeping the overall exploration meaningful.
%{\em Selective symbolic execution} carefully interleaves concrete executions of functions with fully symbolic phases, while keeping the exploration meaningful.

Suppose a function A calls a function B and the execution mode changes at the call site. Two scenarios arise:
(1) {\em From concrete to symbolic and back}: the arguments of B are made symbolic and B is explored symbolically in full. B is also executed concretely and its concrete result is returned to A. After that, A resumes concretely. 
(2) {\em From symbolic to concrete and back}: the arguments of B are concretized, B is executed concretely, and execution resumes symbolically in A. This may impact both soundness and completeness of the analysis: (i) {\em Completeness}: to make sure that symbolic execution skips any paths that would not be realizable due to the performed concretization (possibly leading to false positives), {\sc \stwoe} collects path constraints that keep track of how arguments are concretized, what side effects are made by B, and what return value it produces. (ii) {\em Soundness}: concretization may cause missed branches after A is resumed (possibly leading to false negatives). To remedy this, the collected constraints are marked as {\em soft}: whenever a branch after returning to A is made inoperative by a soft constraint, the execution backtracks and a different choice of arguments for B is attempted. To guide re-concretization of B's arguments, {\sc \stwoe} also collects the branch conditions during the concrete execution of B, and chooses the concrete values so that they enable a different concrete execution path in B.

%===================================================================================
\subsection{Path Selection}
\label{ss:heuristics}

Since enumerating all paths of a program can be prohibitively expensive, in many software engineering activities related to testing and debugging the search is prioritized by looking at the most promising paths first. Among several strategies for selecting the next path to be explored, we now briefly overview some of the most effective ones. %in prior works.
We remark that path selection heuristics are often tailored to help the symbolic engine achieve specific goals (e.g., overflow detection). Finding a universally \mynote{I: universally optimal, ma e' fattibile?}  optimal strategy remains an open problem.

{\em Depth-first search} (DFS), which expands a path as much as possible before backtracking to the deepest unexplored branch, and {\em breadth-first search} (BFS), which expands all paths in parallel, are the most common strategies. DFS is often adopted when memory usage is at a premium, but is hampered by paths containing loops and recursive calls. Hence,  in spite of the higher memory pressure and of the long time required to complete the exploration of specific paths, some tools resort to BFS, which allows the engine to quickly explore diverse paths  detecting interesting behaviors early.
%\iffullver{On the other hand, if the ultimate goal requires to fully terminate the exploration of one or more paths, BFS may take a very long time}{}
Another popular strategy is {\em random path selection}, that has been refined in several variants. For instance, {\sc KLEE}~\cite{KLEE-OSDI08} assigns probabilities to paths based on their length and on the branch arity: it favors paths that have been explored fewer times, preventing starvation caused by loops and other path explosion factors.

% have instead presented
Several works, such as {\sc EXE}~\cite{EXE-CCS06}, {\sc KLEE}~\cite{KLEE-OSDI08}, {\sc Mayhem}~\cite{MAYHEM-SP12}, and {\sc \stwoe}~\cite{CKC-TOCS12}, have discussed heuristics aimed at maximizing code coverage. For instance, the {\em coverage optimize search} discussed in {\sc KLEE}~\cite{KLEE-OSDI08} computes for each state a weight, which is later used to randomly select states. The weight is obtained by considering how far the nearest uncovered instruction is, whether new code was recently covered by the state, and the state's call stack. Of a similar flavor is the heuristic proposed in~\cite{LZL-OOPSLA13}, called {\em subpath-guided search}, which attempts to explore {\it less traveled} parts of a program by selecting the subpath of the control flow graph that has been explored fewer times. This is achieved by maintaining a frequency distribution of explored subpaths, where a subpath is defined as a consecutive subsequence of length $n$ from a complete path. Interestingly, the value $n$ plays a crucial role with respect to the code coverage achieved by a symbolic engine using this heuristic and no specific value has been shown to be universally optimal. %
%Another interesting search strategy is the {\em shortest-distance symbolic execution} heuristic presented in~\cite{MPF-SAS11}. The work does not target coverage, but aims at identifying program inputs that trigger the execution of a specific point in a program. However, similarly to coverage-based strategies, it is based on a metric for evaluating the shortest distance to the target point. This is computed as the length of the shortest path in the inter-procedural control-flow graph, and paths with the shortest distance are prioritized by the engine. 
{\em Shortest-distance symbolic execution}~\cite{MPF-SAS11} does not target coverage, but aims at identifying program inputs that trigger the execution of a specific point in a program. The heuristic is based however, as in coverage-based strategies, on a metric for evaluating the shortest distance to the target point. This is computed as the length of the shortest path in the inter-procedural control flow graph, and paths with the shortest distance are prioritized by the engine. 

%Other search heuristics try to prioritize paths likely leading to states that are {\em interesting} according to some goal. For instance, the {\em buggy-path first} strategy in {\sc AEG}~\cite{AEG-NDSS11} picks paths whose past states have contained small but unexploitable bugs. The intuition is that if a path contains some small errors, it is likely that it has not been properly tested. There is thus a good chance that future states may contain interesting, and hopefully exploitable, bugs. Similarly, the {\em loop exhaustion} strategy discussed in {\sc AEG}~\cite{AEG-NDSS11} explores paths that visit loops. This approach is inspired by the practical observation that common programming mistakes in loops may lead to buffer overflows or other memory-related errors. In order to find exploitable bugs, {\sc Mayhem}~\cite{MAYHEM-SP12} instead gives priority to paths where symbolic memory accesses are identified or symbolic instruction pointers are detected. 

Other search heuristics try to prioritize paths likely leading to states that are {\em interesting} according to some goal. For instance, {\sc AEG}~\cite{AEG-NDSS11} introduces two such strategies. The {\em buggy-path first} strategy picks paths whose past states have contained small but unexploitable bugs. The intuition is that if a path contains some small errors, it is likely that it has not been properly tested. There is thus a good chance that future states may contain interesting, and hopefully exploitable, bugs. Similarly, the {\em loop exhaustion} strategy explores paths that visit loops. This approach is inspired by the practical observation that common programming mistakes in loops may lead to buffer overflows or other memory-related errors. In order to find exploitable bugs, {\sc Mayhem}~\cite{MAYHEM-SP12} instead gives priority to paths where memory accesses to symbolic addresses are identified or symbolic instruction pointers are detected. 

\cite{ZCWDL15} proposes a novel method of dynamic symbolic execution to automatically find a program path satisfying a regular property, i.e., a property (such as file usage or memory safety) that can be represented by a Finite State Machine (FSM). Dynamic symbolic execution is guided by the FSM so that branches of an execution path that are most likely to satisfy the property are explored first. The approach exploits both static and dynamic analysis to compute the priority of a path to be selected for exploration: the states of the FSM that the current execution path has already reached are computed dynamically during the symbolic execution, while backward data-flow analysis is used to compute the future states statically. If the intersection of these two sets is non-empty, there is likely a path satisfying the property.

{\em Fitness functions} have been largely used in the context of search-based test generation~\cite{M-STVR04}. %\mynote{[D] Check and rephrase 2nd sentence}
A fitness function measures how close an explored path is to achieve the target test coverage. Several works, e.g.,~\cite{XTD-DSN09,CS-CACM13}, have applied this idea in the context of symbolic execution. As an example,~\cite{XTD-DSN09} introduces {\em fitnex}, a strategy for flipping branches in concolic execution that prioritizes paths likely {\em closer} to take a specific branch.
In more detail, given a target branch with an associated condition of the form $|a - c| == 0$, the closeness of a path is computed as $|a - c|$ by leveraging the concrete values of variables $a$ and $c$ in that path. Similar fitness values can be computed for other kinds of branch conditions. The path with the lowest fitness value for a branch is selected by the symbolic engine. Paths that have not reached the branch yet get the worst-case fitness value.

\subsection{Symbolic Backward Execution}
\label{ss:backward}
% Symbolic backward execution (SBE)~\cite{CFS-PLDI09,DA-ASE14} is a variant of symbolic execution that proceeds its exploration backward from a target point of a program to an entry point of a program. In other words, the analysis is performed in the reversed direction with respect to a traditional (forward) symbolic execution.
Symbolic backward execution (SBE)~\cite{CFS-PLDI09,DA-ASE14} is a variant of symbolic execution in which the exploration proceeds from a target point to an entry point of a program. The analysis is thus performed in the reverse direction than in canonical (forward) symbolic execution. The main purpose of this approach is typically to identify a test input instance that can trigger the execution of a specific line of code (e.g., an {\tt assert} or {\tt throw} statement). %  
This can be very useful for a developer when performing debugging or regression testing over a program.
As the exploration starts from the target, path constraints are collected along the branches met during the traversal. Multiple paths can be explored at a time by an SBE engine and, akin to forward symbolic execution, paths are periodically checked for feasibility. When a path condition is proved unsatisfiable, the engine discards the path and backtracks.
%SBE starts its exploration from a target point and steps backwards into the code, collecting the path constraints given by the branches that are met during the traversal. In general, multiple paths can be simultaneously explored by a SBE engine. Akin to forward symbolic execution, paths are periodically checked to verify their feasibility. Whenever a path condition is proven to be unsatisfiable, the engine discards the path and then backtracks. 

\cite{MPF-SAS11} discusses a variant of SBE dubbed {\em call-chain backward symbolic execution}~(CCBSE). The technique starts by determining a valid path in the function where the target line is located. When a path is found, the engine moves to one of the callers of the function that contains the target point and tries to reconstruct a valid path from the entry point of the caller to the target point. The process is recursively repeated until a valid path from the main function of the program has been reconstructed. The main difference with respect to the traditional SBE is that, although CCBSE follows the call-chain backwards from the target point, inside each function the exploration is done as in traditional symbolic execution.

%A variant of SBE has been discussed by~\cite{MPF-SAS11} with the {\em call-chain backward symbolic execution} strategy (CCBSE). This technique starts by determining a valid path in the function where the target line is located. When a path is found, the strategy backwards to one of the callers of the function that contains the target point and tries to reconstruct a valid path from the entry point of the caller to the target point. This process is recursively repeated until a valid path from the main function of the program has been reconstructed. The main difference with respect to the traditional SBE is that, although CCBSE follows the call-chain backwards from the target point, inside each function the exploration is done as in traditional symbolic execution.

% TODO
% \iffullver{ (see, e.g., the discussion about CFG reconstruction in Section~\ref{se:symbolic-binary}).}{\mynote{cite appendix or some paper?}.} 
A crucial requirement for the reversed exploration in SBE, as well as in CCBSE, is the availability of the inter-procedural control flow graph which provides a whole-program control flow and makes it possible to determine the call sites for the functions that are involved in the exploration. Unfortunately, constructing such a graph can be quite challenging in practice. Moreover, a function may have many possible call sites, making the exploration performed by a SBE still very expensive. On the other hand, some practical advantages can arise when the constraints are collected in the reverse direction. We will further discuss these benefits in Section~\ref{se:constraint-solving}.

%===================================================================================
\vspace{-2mm}
\subsection{Design Principles of Symbolic Executors}
\label{ss:principles}

A number %\mynote{[D] why MAYHEM?}  
of performance-related design principles that a symbolic execution engine should follow are  summarized in %{\sc Mayhem}
\cite{MAYHEM-SP12}. Most notably:
\begin{enumerate}[topsep=3pt,itemsep=1pt]
  \item {\em Progress}: the executor should be able to proceed for an arbitrarily long time without exceeding the given resources. Memory consumption can be especially critical, due to the potentially gargantuan number of distinct control flow paths.
  \item {\em Work repetition}: no execution work should be repeated, avoiding to restart a program several times from its very beginning in order to analyze different paths that might have a  common prefix.
  \item {\em Analysis reuse}: analysis results from previous runs should be reused as much as possible. In particular, costly invocations to the SMT solver on  previously solved path constraints should be avoided.
\end{enumerate}

\noindent Due to the large size of the execution state space to be analyzed, different symbolic engines have explored different trade-offs between, e.g., running time and memory consumption, or performance and soundness/completeness of the analysis.

Symbolic executors that attempt to execute multiple paths simultaneously in a single run -- also called {\em online} -- clone the execution state at each input-dependent branch. Examples are given in {\sc KLEE}~\cite{KLEE-OSDI08}, {\sc AEG}~\cite{AEG-NDSS11}, {\sc \stwoe}~\cite{CKC-TOCS12}. These engines never re-execute previous instructions, thus avoiding work repetition. However, many active states need to be kept in memory and memory consumption can be large, possibly hindering progress. Effective techniques for reducing the memory footprint include {\em copy-on-write}, which tries to share as much as possible between different states~\cite{KLEE-OSDI08}. As another issue, executing multiple paths in parallel requires to ensure isolation between execution states, e.g., keeping different states of the OS by emulating the effects of system calls.

Reasoning about a single path at a time, as in concolic execution, is the approach taken by so-called {\em offline executors}, such as {\sc SAGE}~\cite{SAGE-NDSS08}. Running each path independently of the others results in low memory consumption with respect to online executors and in the capability of reusing immediately analysis results from previous runs. On the other side, work can be largely repeated, since each run usually restarts the execution of the program from the very beginning. In a typical implementation of offline executors, runs are concrete and require an input seed: the program is first executed concretely, a trace of instructions is recorded, and the recorded trace is then executed symbolically.
{\em Hybrid executors} such as {\sc Mayhem}~\cite{MAYHEM-SP12} attempt at balancing between speed and memory requirements: they start in online mode and generate checkpoints, rather than forking new executors, when memory usage or the number of concurrently active states reaches a threshold. Checkpoints maintain the symbolic execution state and replay information. When a checkpoint is picked for restoration, online exploration is resumed from a restored concrete state.
%: mixed approach. Start with an online approach, if needed switch to offline mode by doing checkpoints. A checkpoint contains the symbolic execution state and replay information. Concrete execution state is discarded since it can be quickly recovered at runtime by using one input generated by the solver before checkpointing.

%% file: memory.tex
% !TEX root = main.tex

%%%%%%%%%%%%%%%%%%%%%%%%%%%%%%%%%%%%%%%%%%%%%%%%%%%%%%%%%%%
\section{Memory model}
\label{memory-model}

Our warm-up example of Section~\ref{symbolic-execution-example} presented a simplified memory model where data are stored in scalar variables only, with no indirection. A crucial aspect of symbolic execution is how memory should be modeled to support programs with pointers and arrays. This requires extending our notion of memory store by mapping not only variables, but also memory addresses to symbolic expressions or concrete values. In general, a store $\sigma$ that explicitly models memory addresses can be thought as a mapping that associates memory addresses (indexes) with either expressions over concrete values or symbolic values. We can still support variables by using their address rather than their name in the mapping. In the following, when we write $x\mapsto e$ for a variable $x$ and an expression $e$ we mean $\&x\mapsto e$, where $\&x$ is the concrete address of variable $x$. Also, if $v$ is an array and $c$ is an integer constant, by $v[c]\mapsto e$ we mean $\&v+c\mapsto e$.

%A memory model is an important design choice for a symbolic engine, as it can have a significant influence on the coverage achieved by symbolic execution, as well as on the scalability of constraint solving~\cite{CS-CACM13}.
\mynote{[D] shorter}A memory model is an important design choice for a symbolic engine, as it can significantly affect the coverage achieved by the exploration and the scalability of constraint solving~\cite{CS-CACM13}.
The {\em symbolic memory address} problem~\cite{SAB-SP10} arises when the address referenced in the operation is a symbolic expression. In the remainder of this section, we discuss a number of popular solutions.

\subsection{Fully Symbolic Memory}
\label{ss:fully-symbolic-memory}

\begin{figure}[t]
%\vspace{-1mm}
\begin{center}
\begin{tabular}{c}
\begin{lstlisting}[basicstyle=\ttfamily\scriptsize]
1.  void foobar(unsigned i, unsigned j) {
2.     int a[2] = { 0 };
3.     if (i>1 || j>1) return;
4.     a[i] = 5;
5.     assert(a[j] != 5);
6.  }
\end{lstlisting}
\end{tabular}
\end{center}
\vspace{-2mm}
\caption{Memory modeling example: which values of \texttt{i} and \texttt{j} make the \texttt{assert} fail?}
\label{fi:example-mem}
\end{figure}

\begin{figure}[t]
%\vspace{-3mm}
\includegraphics[width=1\columnwidth]{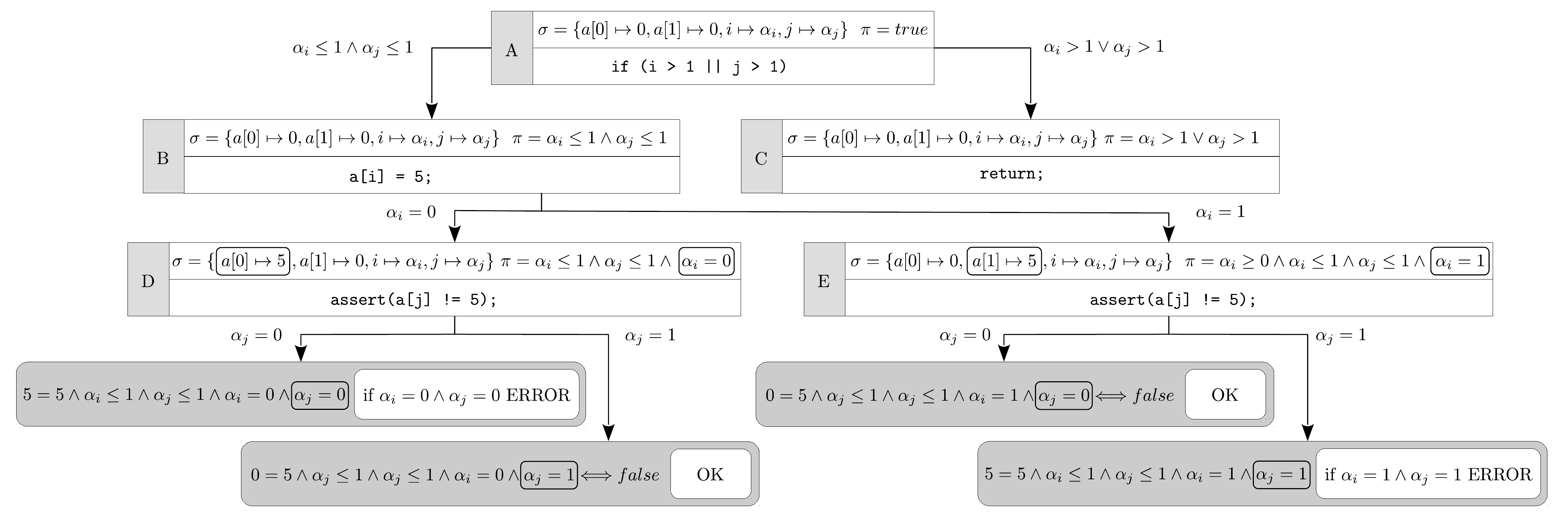} 
\vspace{-4.5mm}
\caption{Fully symbolic memory via state forking for the example of Figure~\ref{fi:example-mem}.}
\label{fi:memory-fork}
%\vspace{-0.5mm}
\end{figure}

At the highest level of generality, an engine may treat memory addresses as fully symbolic. This is the approach taken by a number of works (e.g., {\sc BitBlaze}~\cite{BITBLAZE-ICISS08},~\cite{TLL-CAV10}, {\sc BAP}~\cite{BAP-CAV11}, and~\cite{TS-ATVA14}). Two fundamental approaches, pioneered by King in a seminal paper~\cite{K-CACM76}, are the following:

\begin{itemize}

\item {\em State forking.} If an operation reads from or writes to a symbolic address, the state is forked by considering all possible states that may result from the operation. The path constraints are updated accordingly for each forked state.
\boxedexample{Consider the code shown in Figure~\ref{fi:example-mem}. The write operation at line 4 affects either $a[0]$ or $a[1]$, depending on the unknown value of array index $i$. State forking creates two states after executing the memory assignment to explicitly consider both possible scenarios (Figure~\ref{fi:memory-fork}). The path constraints for the forked states encode the assumption made on the value of $i$. Similarly, the memory read operation \texttt{a[j]} at line 5 may access either $a[0]$ or $a[1]$, depending on the unknown value of array index $j$. Therefore, for each of the two possible outcomes of the assignment \texttt{a[i]=5}, there are two possible outcomes of the \texttt{assert}, which are explicitly explored by forking the corresponding states. }

\begin{figure}[t]
\begin{center}
\includegraphics[width=0.75\columnwidth]{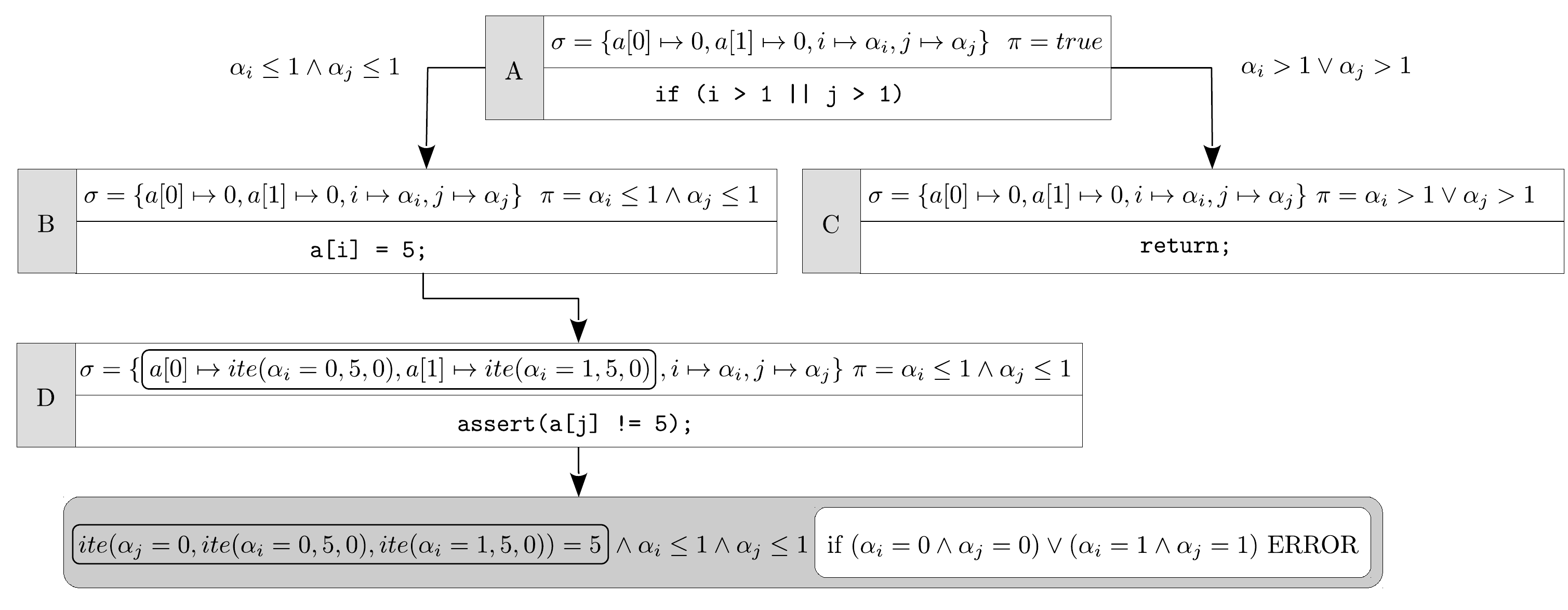}
\end{center}
\vspace{-3mm}
\caption{Fully symbolic memory via if-then-else formulas for the example of Figure~\ref{fi:example-mem}.}
%\vspace{-1mm} % TODO
\label{fi:memory-ite}
\vspace{-1.5mm}
\end{figure}

% otherwise\footnote{In propositional logic, the $ite(\texttt{c}, \texttt{t}, \texttt{f})$ expression could be replaced with the formula $(\texttt{c} \wedge \texttt{t}) \vee (\neg\texttt{c} \wedge \texttt{f})$.}.
\item {\em if-then-else formulas.} An alternative approach consists in encoding the uncertainty on the possible values of a symbolic pointer into the expressions kept in the symbolic store and in the path constraints, without forking any new states. The key idea is to exploit the capability of some solvers to reason on formulas that contain if-then-else expressions of the form $ite(\texttt{c}, \texttt{t}, \texttt{f})$, which yields \texttt{t} if \texttt{c} is true, and \texttt{f} otherwise.
The approach works differently for memory read and write operations. Let $\alpha$ be a symbolic address that may assume the concrete values $a_1, a_2, \ldots$:
\begin{itemize}
\item reading from $\alpha$ yields the expression $ite(\alpha=a_1,\sigma(a_1), ite(\alpha=a_2,\sigma(a_2), \ldots))$;
\item writing an expression $e$ at $\alpha$ updates the symbolic store for each $a_1, a_2, \ldots$ as $\sigma(a_i)\gets ite(\alpha=a_i,e,\sigma(a_i))$.
\end{itemize}
Notice that in both cases, a memory operation introduces in the store as many $ite$ expressions as the number of possible values the accessed symbolic address may assume. The $ite$ approach to symbolic memory is used, e.g., in {\sc Angr}~\cite{ANGR-SSP16} (Section~\ref{ss:index-based-memory}).
\boxedexample{Consider again the example shown in Figure~\ref{fi:example-mem}. Rather than forking the state after the operation \texttt{a[i]=5} at line 4, the if-then-else approach updates the memory store by encoding both possible outcomes of the assignment, i.e., $a[0]\mapsto ite(\alpha_i=0,5,0)$ and $a[1]\mapsto ite(\alpha_i=1,5,0)$ (Figure~\ref{fi:memory-ite}). Similarly, rather than creating a new state for each possible distinct address of \texttt{a[j]} at line 5, the uncertainty on $j$ is encoded in the single expression $ite(\alpha_j=0,\sigma(a[0]),\sigma(a[1]))=ite(\alpha_j=0,ite(\alpha_i=0,5,0),ite(\alpha_i=1,5,0))$.
%: if $\alpha_i=0$ then $a[0]\mapsto 5$ and $a[1]\mapsto 0$; conversely, if $\alpha_i=1$ then $a[0]\mapsto 0$ and $a[1]\mapsto 5$.
%State forking creates two states after executing the memory assigment to explicitly consider both possible scenarios (Figure~\ref{fi:memory-fork}). The path constraints for the forked states encode the assumption made on the value of $i$. Similarly, the memory read operation \texttt{a[j]} at line 5 may access either $a[0]$ or $a[1]$, depending on the unknown value of array index $j$. Therefore, for each of the two possible outcomes of the assignment \texttt{a[i]=5}, there are two possible outcomes of the \texttt{assert}, which are explicitly explored by forking the corresponding states. 
}

%Indeed, the $ite(\texttt{c}, \texttt{t}, \texttt{f})$ expression introduced in the symbolic store $\sigma$ is a short term for an {\tt if-then-else} expression and means that if the condition {\tt c} is verified then {\tt t} holds, otherwise {\tt f} must be assumed as true. Nonetheless, $ite$ expressions are often just syntactic sugar for disjunctive formulas and are commonly supported by most prominent constraint solvers. For instance, in the context of propositional logic the $ite(\texttt{c}, \texttt{t}, \texttt{f})$  expression could be replaced with the formula $(\texttt{c} \wedge \texttt{t}) \vee (\neg\texttt{c} \wedge \texttt{f})$ . 

\end{itemize}

%\noindent To model fully symbolic pointers, an extensive line of research (e.g., {\sc EXE}~\cite{EXE-CCS06}, {\sc KLEE}~\cite{KLEE-OSDI08}, {\sc SAGE}~\cite{EGL-ISSTA09}) leverages the expressive power of SMT solvers to model array operations as first-class entities in constraint formulas using a {\em theory of arrays} in the decision procedure~\cite{STP-CAV07}.

%\noindent % TODO trick if you need one more line
An extensive line of research (e.g., {\sc EXE}~\cite{EXE-CCS06}, {\sc KLEE}~\cite{KLEE-OSDI08}, {\sc SAGE}~\cite{EGL-ISSTA09}) leverages the expressive power of some SMT solvers to model fully symbolic pointers. Using a {\em theory of arrays}~\cite{STP-CAV07}, array operations can in fact be expressed as first-class entities in constraint formulas.

Due to its generality, fully symbolic memory supports the most accurate description of the memory behavior of a program, accounting for all possible memory manipulations. In many practical scenarios, the set of possible addresses a memory operation may reference is small~\cite{BITBLAZE-ICISS08} as in the example shown in Figure~\ref{fi:example-mem} where indexes $i$ and $j$ range in a bounded interval, allowing accurate analyses using a reasonable amount of resources. In general, however, a symbolic address may reference any cell in memory, leading to an intractable explosion in the number of possible states. For this reason, a number of techniques have been designed to improve scalability, which elaborate along the following main lines:

\begin{itemize}
\item {\em Representing memory in a compact form.} This approach was taken in~\cite{MEMSIGHT-ASE17}, which maps symbolic -- rather than concrete -- address expressions to data, representing the possible alternative states resulting from referencing memory using symbolic addresses in a compact, implicit form. Queries are offloaded to efficient paged interval tree implementations to determine which stored data are possibly referenced by a memory read operation.

\item {\em Trading soundness for performance.} The idea, discussed in the remainder of this section, consists in corseting symbolic exploration to a subset of the execution states by replacing symbolic pointers with concrete addresses.

\item {\em Heap modeling.} An additional idea is to corset the exploration to states where pointers are restricted to be either null, or point to previously heap-allocated objects, rather than to any generic memory location (Section~\ref{ss:address-concretization} and Section~\ref{ss:complex-objects}).
\end{itemize}

%When obtained ranges are too large, {\sc BitBlaze}~\cite{BITBLAZE-ICISS08} adds a further constraint to the system to limit its size. However, the authors observe that most symbolic memory accesses are typically already constrained to small ranges in practice, making it unnecessary.

%\vspace{-2pt} % TODO
\subsection{Address Concretization}
\label{ss:address-concretization}

In all cases where the combinatorial complexity of the analysis explodes as pointer values cannot be bounded to sufficiently small ranges, {\em address concretization}, which consists in concretizing a pointer to a single specific address, is a popular alternative. This can reduce the number of states and the complexity of the formulas fed to the solver and thus improve running time, although may cause the engine to miss paths that, for instance, depend on specific values for some pointers.

%Systems such as {\sc CUTE}~\cite{CUTE-FSE05} and {\sc CREST}~\cite{CREST-ASE08} are capable of reasoning only about equality constraints for pointers, as they can be solved efficiently, and resort to concretization for general symbolic references. % equality and inequality

%\mynote{DART is mentioned in CS-CACM13 as  using theories of arrays} --> added to the list above.
Concretization naturally arises in offline executors (Section~\ref{ss:principles}). Prominent examples are {\sc DART}~\cite{DART-PLDI05} and {\sc CUTE}~\cite{CUTE-FSE05},
%and early {\sc SAGE} releases~\cite{SAGE-NDSS08}. % that concretely execute one path at a time while collecting path constraints along executed paths. %\mynote{[D] was: equality and inequality}  
which handle memory initialization by concretizing a reference of type {\tt T*} either to {\tt NULL}, or to the address of a newly allocated object of {\tt sizeof(T)} bytes. DART makes the choice randomly, while CUTE first tries {\tt NULL}, and then, in a subsequent execution, a concrete address. If {\tt T} is a structure, the same concretization approach is recursively applied to all fields of a pointed object. Since memory addresses (e.g., returned by {\tt malloc}) may non-deterministically change at different concrete executions, CUTE uses {\em logical addresses} in symbolic formulas to maintain consistency across different runs.
Another reason for concretization is due to efficiency in constraint solving: for instance, CUTE reasons only about pointer equality constraints using an equivalence graph, resorting to concretization for more general constraints that would need costly SMT theories.
\subsection{Partial Memory Modeling}
\label{ss:index-based-memory}

To mitigate the scalability problems of fully symbolic memory and the loss of soundness of memory concretization,
%Motivated by the observation that concretizing all memory indexes might not work well in some scenarios, while fully symbolic memory does not scale, 
{\sc Mayhem}~\cite{MAYHEM-SP12} explores a middle point in the spectrum by introducing a {\em partial} memory model. The key idea is that written addresses are always concretized and read addresses are modeled symbolically if the contiguous interval of possible values they may assume is small enough. This model is based on a trade-off: it uses more expressive formulas than concretization, since it encodes multiple pointer values per state, but does not attempt to encode all of them like in fully symbolic memory~\cite{MAYHEM-THESIS}. A basic approach to bound the set of possible values that an address may assume consists in trying different concrete values and checking whether they satisfy the current path constraints, excluding large portions of the address space at each trial until a tight range is found.  
%This choice is important to keep the analysis feasible: for instance, in a fully symbolic model a repeated read and write on the same symbolic index would result in quadratic increase in either the symbolic constraints or the complexity of the stored symbolic expressions~\cite{DRILLER-NDSS16}.
%Global memory is defined as a map $\mu$ from 32-bit addresses ({\em indexes}) to expressions. When a symbolic index $i$ is used to read memory, the algorithm generates a memory object $M$ containing the projection of $\mu$ over all the valid values that $i$ can assume. The evaluation of a $load(\mu,i)$ operation is thus reduced to $M[i]$, where $M$ is typically orders of magnitude smaller than the entire memory $\mu$.
%Instantiating a memory object still requires finding all the possible values for a symbolic index. A naive algorithm would employ the constraint solver to refine the range of an index using binary search under the current path constraints. 
This algorithm comes with a number of caveats: for instance, querying the solver on each symbolic dereference is expensive, the memory range may not be continuous, and the values within the memory region of a symbolic pointer might have structure. {\sc Mayhem} thus performs a number of optimizations such as {\em value-set analysis}~\cite{VSA-CC04} and forms of query caching (Section~\ref{se:constraint-solving}) to refine ranges efficiently. If at the end of the process the range size exceeds a given threshold (e.g., 1024), the address is concretized. {\sc Angr}~\cite{ANGR-SSP16} also adopts the partial memory model idea and extends it by optionally supporting write operations on symbolic pointers that range within small contiguous intervals (up to 128 addresses). % [D] ptr may also be redirected to symbolic data

%%%%%%%%%%%%%%%%%%%%%%%%%%%%%%%%%%%%%%%%%%%%%%%%%%%%
%\subsection{Complex Objects}
%

\subsection{Lazy Initialization}
\label{ss:complex-objects}

\cite{KPV-TACAS03} proposes symbolic execution techniques for advanced object-oriented language constructs, such as those offered by C++ and Java. The authors describe a framework for software verification that combines symbolic execution and model checking to handle linked data structures such as lists and trees. % [D] added dynamically allocated & discarded primitive data types, and concurrency.

In particular, they generalize symbolic execution by introducing {\em lazy initialization} to effectively handle dynamically allocated objects. Compared to our warm-up example from Section~\ref{symbolic-execution-example}, the state representation is extended with a {\em heap configuration} used to maintain such objects. Symbolic execution of a method taking complex objects as inputs starts with uninitialized fields, and assigns values to them in a lazy fashion, i.e., they are initialized when first accessed during execution.

When an uninitialized reference field is accessed, the algorithm forks the current state with three different heap configurations, in which the field is initialized with: (1) {\tt null}, (2) a reference to a new object with all symbolic attributes, and (3) a previously introduced concrete object of the desired type, respectively. \iffullver{This on-demand concretization enables symbolic execution of methods without the need for any previous knowledge on the number of objects given as input. Also, forking the state as in (2) results into a systematic treatment for aliasing, i.e., when an object can be accessed through multiple references.}{}

\cite{KPV-TACAS03,SPF-ISSTA04} combine lazy initialization with user-provided {\em method preconditions}, i.e., conditions that are assumed to be true before the execution of a method. Preconditions are used to characterize those program input states in which the method is expected to behave as intended by the programmer. For instance, we expect a binary tree data structure to be acyclic and with every node - except for the root - having exactly one parent. Conservative preconditions are used to ensure that incorrect heap configurations are eliminated during initialization, speeding up the symbolic execution process. %\mytempedit{To better illustrate this technique, we now discuss an example in which lazy initialization is used to handle a {\tt struct} data type.}

\begin{figure*}[t]
  %\vspace{-3mm}
  \centering
  \includegraphics[width=0.875\columnwidth]{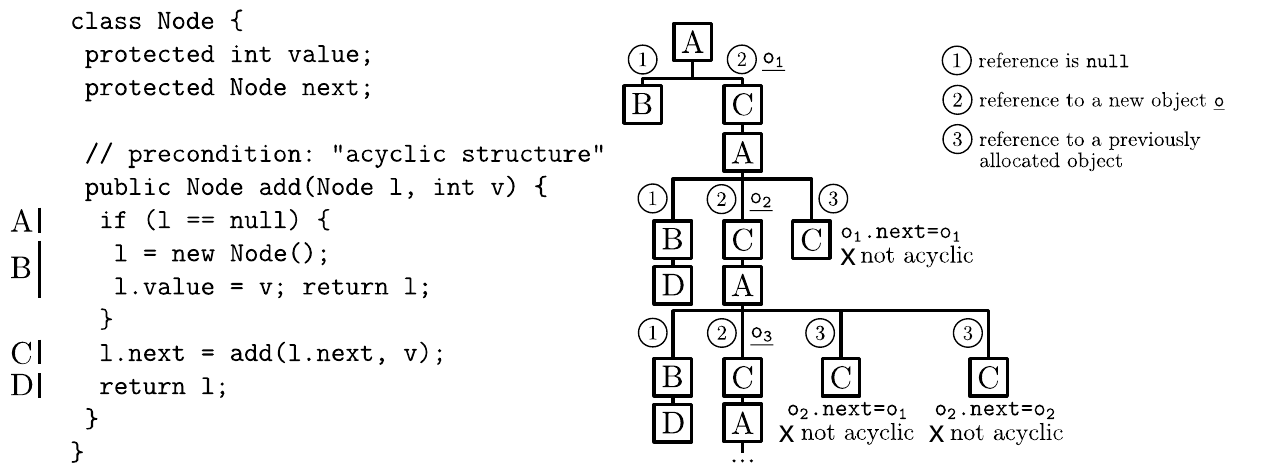} % TODO was 0.9
  \vspace{-0.75mm}
  \caption{Example of lazy initialization}
  \label{fig:example-lazy-initialization}
  %\vspace{-3mm}
\end{figure*}

\boxedexample{
% For the sake of simplicity, we assume that fragment C does not actually evaluate {\tt l->next}, but leaves this task to fragment A. When expanding the [...] 
%the value of
 Figure~\ref{fig:example-lazy-initialization} shows a recursive Java method {\tt add}, which appends a node of type {\tt Node} to a linked list, and a minimal representation of its symbolic execution when applying lazy initialization. The tree nodes represent executions of straight-line fragments of {\tt add}. Initially, fragment A evaluates reference {\tt l}, which is symbolic and thus uninitialized. The symbolic engine considers three  options: (1) {\tt l} is {\tt null}, (2) {\tt l} points to a new object, and (3) {\tt l} points to a previously allocated object. Since this is the first time that a reference of type {\tt Node} is met, option (3) is ruled out. The two remaining options are then expanded, executing the involved fragments. While the first path ends after executing fragment B, the second one implicitly creates a new object {\tt o$_\texttt{1}$} due to lazy initialization and then executes C, recursively invoking {\tt add}. When expanding the recursive call, fragment A is executed and the three options are again considered by the engine, which forks into three distinct paths. Option (3) is now taken into account since a {\tt Node} object has been previously allocated (i.e., {\tt o$_\texttt{1}$}). However, this path is soon aborted by the engine since it violates the acyclicity precondition (expressed as a comment in this example). The other forked paths are further expanded, repeating the same process. Since the linked list has an unknown maximum length, the exploration can proceed indefinitely. For this reason, it is common to assume an upper bound on the depth of the materialization (i.e., field instantiation) chain.
}

Recent advances in the area have focused on improving efficiency in generating heap configurations. For instance, in~\cite{DLR-ASE12} the concretization of a reference variable is deferred until the object is actually accessed. The work also provides a formalization of lazy initialization. \cite{BLISS-TSE15} instead employs bound refinement to prune uninteresting heap configurations by using information from already concretized fields, while a SAT solver is used to check whether declarative -- rather than imperative as in the original algorithm -- preconditions hold for a given configuration.

\iffullver{
\myparagraph{Verifying Client Code Only}
Of a different flavor is the technique presented in~\cite{SHZ-TAIC07} for symbolic execution over objects instantiated from commonly used libraries. The authors argue that performing symbolic execution at the representation level might be redundant if the aim is to only check the client code, thus trusting the correctness of the library implementation. They discuss the idea of symbolically executing methods of the Java {\tt String} class using a finite-state automaton that abstracts away the implementation details. They present a case study of an application that dynamically generates SQL queries: symbolic execution is used to check whether the statements conform to the SQL grammar and possibly match injection patterns. \iffullver{The authors mention that their approach might be used to symbolically execute over standard container classes such as trees or maps. It is worth mentioning that symbolic execution is used to detect SQL injection vulnerabilities also in~\cite{FLP-COMPSAC07}.}{The authors mention that their approach might be used to symbolically execute over standard container classes such as trees or maps.}
}{}

%% file: environment.tex
\section{Interaction with the environment}
\label{se:environment-thirdparty}

As most programs are not self-contained, a symbolic engine has to take into account their frequent interactions with the surrounding software stack. A typical example is data flows that take place through features of the underlying operating system (e.g., file system, environment variables, network). Functions controlled by the system environment are often referred to as {\em external}. Modern applications pose further challenges when they interact with the user via other components (e.g., Swing, Android), or invoke special features of their execution runtimes. Missing symbolic data flows through these software elements might indeed affect the meaningfulness of the analysis.
 
\myparagraph{System Environment}
% symbolic analysis \revedit
A body of early works (e.g., {\sc DART}~\cite{DART-PLDI05},  {\sc CUTE}~\cite{CUTE-FSE05}, and {\sc EXE}~\cite{EXE-CCS06}) include the system environment in the analysis by actually executing external calls using concrete arguments for them. This indeed limits the behaviors they can explore compared to a fully symbolic strategy, which on the other hand might be unfeasible. 
In an online executor this choice may also result in having external calls from distinct paths of execution interfere with each other. As there is no mechanism for tracking the side effects of each external call, there is potentially a risk of state inconsistency, e.g., an execution path may read from a file while at the same time another execution path is trying to delete it.

%Another way to tackle the problem is to create an abstract model that captures these interactions 
A way to overcome this problem is to create abstract models that capture these interactions. For instance, in {\sc KLEE}~\cite{KLEE-OSDI08} symbolic files are supported through a basic {\em symbolic file system} for each execution state, consisting of a directory with $n$ symbolic files whose number and sizes are specified by the user. An operation on a symbolic file results in forking $n+1$ state branches: one for each possible file, plus an optional one to capture unexpected errors in the operation. As the number of functions in a standard library is typically large and writing models for them is expensive and error-prone~\cite{Ball06}, models are generally implemented at system call-level rather than library level. This enables the symbolic exploration of the libraries as well.

% {\sc CLOUD9} further extends support to many other POSIX libraries, allowing users to also control advanced conditions in the testing environment. For instance, it can simulate reordering, delays, and packet dropping caused by a fragmented data stream over a network.
{\sc AEG}~\cite{AEG-NDSS11} models most of the system environment that could be used by an attacker as input source, including the file system, network sockets, and environment variables. Additionally, more than 70 library and system calls are emulated, including thread- and process-related system calls, and common formatting functions to capture potential buffer overflows. Symbolic files are handled as in {\sc KLEE}~\cite{KLEE-OSDI08}, while symbolic sockets are dealt with in a similar manner, with packets and their payloads being processed as in symbolic files and their contents. {\sc CLOUD9}~\cite{CLOUD9-EUROSYS11} supports additional POSIX libraries, and allows users to control advanced conditions in the testing environment. For instance, it can simulate reordering, delays, and packet dropping caused by a fragmented network data stream.

{\sc \stwoe}~\cite{CKC-TOCS12} remarks that models, other than expensive to write, rarely achieve full accuracy, and may quickly become stale if the modeled system changes. 
%\sc \stwoe}~\cite{CKC-TOCS12} remarks that models, other than expensive to write, rarely achieve full accuracy, and may quickly become stale if the modeled system changes. Hence, it would be preferable to let analyzed programs interact with the real environment while exploring multiple paths. In their \stwoe\ platform, the authors rely on virtualization to perform the desired analysis on the real software stack, preventing side effects from propagating across independent execution paths.
It would thus be preferable to let analyzed programs interact with the real environment while exploring multiple paths. However, this must be done without incurring in environment interferences or state inconsistencies. To achieve this goal, \stwoe\ resorts to virtualization to prevent propagation of side effects across independent execution paths when interacting with the real environment. QEMU is used to emulate the full software stack: instructions are transparently translated into micro operations run by the native host, while an x86-to-LLVM lifter is used to perform symbolic execution of the instructions sequence in {\sc KLEE}~\cite{KLEE-OSDI08}. This allows {\sc \stwoe} to properly evaluate any side effects due to the environment. Notice that whenever a symbolic branch condition is evaluated, the execution engine forks a parallel instance of the emulator to explore the alternative path. Selective symbolic execution (Section~\ref{ss:concrete-concolic-symbolic}) is used to limit the scope of symbolic exploration across the software stack, partially mitigating the overhead of emulating a full stack (e.g., user code, libraries, drivers) that can significantly limit the scalability of the overall solution. % QEMU~\cite{QEMU-ATC05}

{\sc DART}'s approach~\cite{DART-PLDI05} is different, as the goal is to enable automated unit testing. DART deems as foreign interfaces all the external variables and functions referenced in a C program along with the arguments for a top-level function.
%External interfaces are identified by parsing the source code, and a random test driver is generated accordingly.
External functions are simulated by nondeterministically returning any value of their specified return type. To allow the symbolic exploration of library functions that do not depend on the environment, the user can adjust the boundary between external and non-external functions to tune the scope of the symbolic analysis.
%Library functions are normally not considered external functions as they are controlled by the program, but in practice the user can adjust the boundary between library and external functions to simulate the desired effects.

% symbolically analyzed
\myparagraph{Application Environment}
We now discuss possible solutions for dealing with software elements that carry out control and data flows on the behalf of the program under analysis. Instances of this problem arise for instance in frameworks like Swing and Android, which embody abstract designs to invoke application code (e.g., via callbacks) during user interaction~\cite{JQF-ICSE16}. Symbolic values flow outside the boundaries of the analysis also for applications running in managed runtimes, e.g., when calling native Java methods or unmanaged code in .NET~\cite{AOH-TACAS07}. Such features complicate the implementation of an engine: for instance, native methods and reflection in Java depend on the internals of the underlying JVM~\cite{Anand12}. Closed-source components might represent another instance of this problem.

% (previous template) we should cite EXE-TECH06 but we get the same identifier :-(
%{\sc EXE}~\cite{EXE-CCS06} instead devises a trial-and-error strategy so that calls with symbolic arguments to external library functions are logged, and the user has to manually intervene to instrument the external code and restart the exploration. This methodology aims at ensuring more complete constraint generation compared to concretization, but may require repeated interaction with the user and not scale for large functions.

%possibly resulting in an incomplete exploration and in turn failing to generate test inputs for feasible program paths
Similarly as in system environment modeling, early works such as {\sc DART}~\cite{DART-PLDI05} and {\sc CUTE}~\cite{CUTE-FSE05} deal with calls to other software components by executing them with concrete arguments. This may result in an incomplete exploration, failing to generate test inputs for feasible program paths. On the other hand, a symbolic execution of their code is unlikely to succeed for a number of reasons: for instance, the implementation of externally simple behaviors is often complex as it has to allow for extensibility and maintainability, or may contain details irrelevant to the exploration, such as how to display a button that triggers a callback~\cite{JQF-ICSE16}.
%
%such components often devise complex implementations of externally simple behaviors in order to allow for extensibility and maintainability, or may contain details irrelevant to the exploration such as the graphical representation of a feature.
%
One solution would be to mimic external components with simpler and more abstract models. However, writing component models manually -- which can be a daunting task per se -- might be hard due to the unavailability of the source code, and applications using unsupported models would remain out of reach.

%as in general models cannot be reused across components, applications that use unsupported ones remain out of reach. 

% that are heavily used
Some works (e.g., \cite{AOH-TACAS07,XXT-ICSE11}) explore techniques to pinpoint which entities from a component may hold symbolic values in a symbolic exploration, and thus require human intervention (e.g., writing a model) for their analysis. A different line of research has instead attempted to generate models automatically, which may be the only viable option for closed-source components. \cite{CT-SEN14,VTV-SEN15} employ program slicing to extract the code that manipulates a given set of fields relevant for the analysis, and build abstract models from it. \cite{JQF-ICSE16} takes a step further by using program synthesis to produce models for Java frameworks. Such models provide equivalent instantiations of design patterns that are heavily used in many frameworks: this helps symbolic executors discover control flow -- such as callbacks to user code through an observer pattern -- that would otherwise be missed. An advantage of using program synthesis is that it can generate more concise models than slicing, as it abstracts away the details and entanglements of how a program is written by capturing its functional behavior.

%% file: explosion.tex
\section{Path explosion}
\label{se:path-explosion}

One of the main challenges of symbolic execution is the path explosion problem: a symbolic executor may fork off a new state at every branch of the program, and the total number of states may easily become exponential in the number of branches. Keeping track of a large number of pending branches to be explored, in turn, impacts both the running time and the space requirements of the symbolic executor. 

The main sources of path explosion are loops and function calls. Each iteration of a loop can be seen as an {\tt if-goto} statement, leading to a conditional branch in the execution tree. If the loop condition involves one or more symbolic values, the number of generated branches may be potentially infinite, as suggested by the following example. 

%\vspace{-3pt}
\begin{shaded}\noindent{\bf\small Example.} Consider the following code fragment~\cite{CS-CACM13}:
  \vspace{-1mm}
  \begin{lstlisting}[basicstyle=\ttfamily\small]
                int x = sym_input(); // e.g., read from file
                while (x > 0) x = sym_input();  
  \end{lstlisting}
  \vspace{-1mm}
\noindent where \texttt{sym\_input()} is an external routine that interacts with the environment (e.g., by reading input data from a network) and returns a fresh symbolic input. The path constraint set at any final state has the form:
\vspace{-1mm}
\[ \pi = \left( {\bigwedge}_{i \in [1, k]}\,\alpha_i > 0 \right) \wedge (\alpha_{k+1} \leq 0) \]
\noindent where $k$ is the number of iterations and $\alpha_i$ is the symbol produced by \texttt{sym\_input()} at the $i$-th iteration.
\end{shaded}
%\vspace{-3pt} 

%\noindent
While it would be simple (and is indeed common) to bound the loop exploration up to a limited number of iterations, interesting paths could be easily missed with this approach.  A large number of works have thus explored more advanced strategies, e.g., by characterizing similarities across distinct loop iterations or function invocations through summarization strategies that prevent repeated explorations of a code portion or by inferring invariants that inductively describe the properties of a computation. In the remainder of this section we present a variety of prominent techniques, often based on the computation of an under-approximation of the analysis with the aim of exploring only a relevant subset of the state space.

% ---------------------------------------------------------------------------------------------------
\subsection{Pruning Unrealizable Paths}
\label{ss:unrealizable-paths}

A first natural strategy to reduce the path space is to invoke the constraint solver at each branch, pruning unrealizable branches: if the solver can prove that the logical formula given by the path constraints of a branch is not satisfiable, then no assignment of the program input values could drive a real execution toward that path, which can be safely discarded by the symbolic engine without affecting soundness. An example of this strategy is provided in Figure~\ref{fig:eager-evaluation}.

\iffullver{
\boxedexample{Consider the example shown in Figure~\ref{fig:eager-evaluation} and assume that {\tt a} is a local variable bound to an unconstrained symbol $\alpha_a$. A symbolic engine would start the execution of the code fragment of Figure~\ref{fig:eager-evaluation}a by evaluating the branch condition $a > 0$. Before expanding both branches, the symbolic engine queries a constraint solver to verify that no contradiction arises when adding to the path constraints $\pi$ the {\em true} branch condition ($\alpha_a > 0$) or the {\em false} branch condition ($\alpha_a \leq 0$). Since both paths are feasible, the engine forks the execution states B and D (see Figure~\ref{fig:eager-evaluation}b). A similar scenario happens when the engine evaluates the branch condition $a > 1$. However, since $\alpha_a$ is not unconstrained anymore, some contradictions are actually possible. The engine queries the solver to check the following path constraints: (1) $\alpha_a > 0~\wedge~\alpha_a > 1$, (2)~$\alpha_a > 0~\wedge~\alpha_a \leq 1$, (3) $\alpha_a \leq 0~\wedge~\alpha_a > 1$, and (4) $\alpha_a \leq 0~\wedge~\alpha_a \leq 1$. The formula $\alpha_a \leq 0 \wedge \alpha_a > 1$, however, does not admit a valid solution and therefore the related path can be safely dropped by the engine. %On the other hand, other paths admit a valid solution and can be further explored by the engine.
}
}

\begin{figure}[t]
  %\vspace{-3mm}
  %\centering
  \begin{subfigure}{.29\textwidth}
    %\vspace{0mm}
    \begin{lstlisting}[basicstyle=\ttfamily\scriptsize]
   if (a > 0) { ... } 

   if (a > 1) { ... }
    \end{lstlisting}
    \vspace{4.99mm}
    \caption{}
  \end{subfigure}%
  %\hspace{-2mm}
  \begin{subfigure}{.70\textwidth}
    \centering
    \includegraphics[width=1.0\columnwidth]{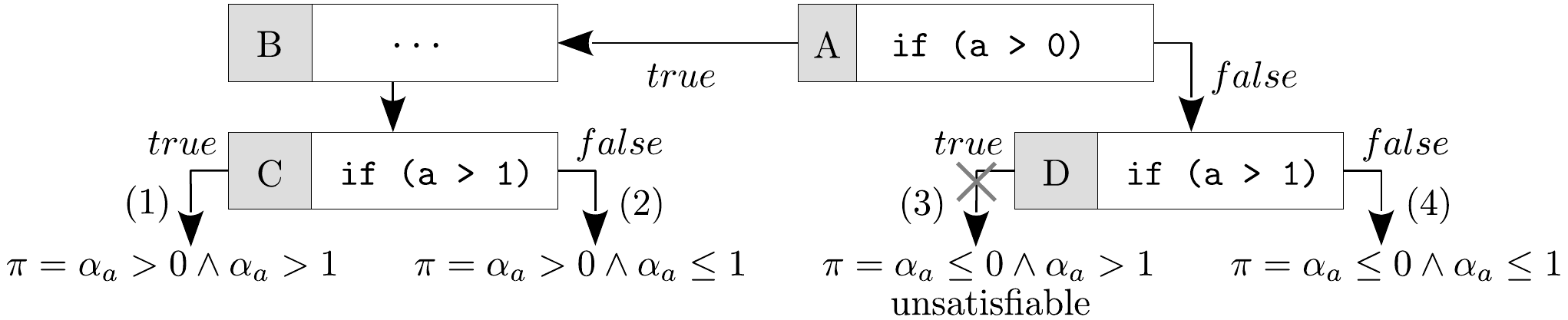} 
    %\label{fig:sub1}
    \vspace{-6mm}
    \caption{}
  \end{subfigure}%
  \vspace{-2mm}
  \caption{Pruning unrealizable paths example: (a) code fragment; (b) symbolic execution of the code fragment: the {\em true} branch at node D is not explored since its path constraints $(\alpha_a \leq 0 \wedge \alpha_a > 1)$ are not satisfiable.}
  \label{fig:eager-evaluation}
\end{figure}

This approach is commonly referred to as {\em eager evaluation} of path constraints, since constraints are eagerly checked at each branch, and is typically the default in most symbolic engines. We refer to Section~\ref{se:constraint-solving} for a discussion of the opposite strategy, called {\em lazy evaluation}, aimed at reducing the burden on the constraint solver.

% This can can be seen as a set of learned conflicts and 
An orthogonal approach that can help reduce the number of paths to check is presented in~\cite{SSJ-NFM15}. While an SMT solver can be used to explore a large search space one path at a time, it will often end up reasoning over control flows shared by many paths. The work exploits this observation by extracting a minimal {\em unsat core} from each path that is proved to be unsatisfiable,
removing as many statements as possible while preserving unsatisfiability. An engine could thus exploit unsat cores to discard paths that share the same (unsatisfiable) statements.

\subsection{Function and Loop Summarization} 
\label{ss:summarization}

When a code fragment -- be it a function or a loop body --  is traversed several times, the symbolic executor can build a summary of its execution for subsequent reuse.

\myparagraph{Function Summaries}
A function $f$ may be called multiple times throughout the execution, either at the same calling context or at different ones. Differently from plain executors, which would execute $f$ symbolically at each invocation, the compositional approach proposed in~\cite{G-POPL07} for concolic executors dynamically generates {\em function summaries}, allowing the executor to effectively reuse prior discovered analysis results. The technique captures the effects of a function invocation with a formula $\phi_w$ that conjoins constraints on the function inputs observed during the exploration of a path $w$, describing equivalence classes of concrete executions, with constraints observed on the outputs. Inputs and outputs are defined in terms of accessed memory locations.
A function summary is a propositional logic formula defined as the disjunction of $\phi_w$ formulas from distinct classes, and feasible inter-procedural paths are modeled by composing symbolic executions of intra-procedural ones.
\cite{AGT-TACAS08} extends compositional symbolic execution by generating summaries as first-order logic formulas with uninterpreted functions, allowing the formation of incomplete summaries (i.e., capturing only a subset of the paths within a function) that can be expanded on demand during the inter-procedural analysis as more statements get covered. 
% skipping it for now: not self-contained
%\cite{CFS-PLDI09} makes a step further by describing an algorithm for generalizing symbolic summaries, effectively enhancing their reuse in practice.

%A similar idea has been also proposed in~\cite{BCE-TACAS08}. The main intuition is that,
\cite{BCE-TACAS08} explores a different flavor of summarization, based on the following intuition: if two states differ only for some program values that are not read later, the executions generated by the two states will produce the same side effects. Side effects of a code fragment can be therefore cached and possibly reused later.

\myparagraph{Loop Summaries}
Akin to function calls, partial summarizations for loops can be obtained as described in~\cite{GL-ISSTA11}. A loop summary uses pre- and post-conditions that are dynamically computed  during the symbolic execution by reasoning on the dependencies among loop conditions and symbolic variables. Caching loop summaries not only allows the symbolic engine to avoid redundant executions of the same loop in the same program state, but  makes it also possible to generalize the summary to cover different executions of the same loop under different conditions. 

Early works can generate summaries only for loops that update symbolic variables across iterations by adding a fixed amount to them. Also, they cannot handle nested loops or {\em multi-path loops}, i.e., loops with branches within their body. Proteus~\cite{XX-FSE16} is a general framework proposed for summarizing multi-path loops. It classifies loops according to the patterns of values changes in path conditions (i.e., whether an induction variable is updated) and of the interleaving of paths within the loop (i.e., whether there is a regularity). The classification leverages an extended form of control flow graph, which is then used to construct an automata that models the interleaving. The automata is traversed in a depth-first fashion and a disjunctive summarization is constructed for all the feasible traces in it, where a trace represents an execution in the loop. The classification determines if a loop can be captured either precisely or approximately (which can still be of practical relevance), or it cannot. Precise summarization of multi-path loops with irregular patterns or non-inductive updates, and more importantly summarization of nested loops remain open research problems.

Of a different flavor is the compaction technique introduced in~\cite{SST-ATVA13}, where\mynote{I: Ha senso qui? Altrimenti dove?} the analysis of cyclic paths in the control flow graph yields {\em templates} that declaratively describe the program states generated by a portion of code as a {\em compact} symbolic execution tree. By exploiting templates, the symbolic execution engine can explore a significantly reduced number of program states. A drawback of this approach is that templates introduce quantifiers in the path constraints: in turn, this may significantly increase the burden on the constraint solver.

% ---------------------------------------------------------------------------------------------------
\subsection{Path Subsumption and Equivalence} 
\label{ss:interpolation}

A large symbolic state space offers scope for techniques that explore path similarity to, e.g., discard paths that cannot lead to new findings, or abstract away differences when profitable. In this section we discuss a number of works along these lines.

\myparagraph{Interpolation}
Modern SAT solvers rely on a mutual reinforcing combination of search and deduction, using the latter to drive the former away from a conflict when it becomes blocked. In a similar manner, symbolic execution can benefit from {\em interpolation} techniques to derive properties from program paths that did not show a desired property, so to prevent the exploration of similar paths that would not satisfy it either.
%so to prevent the execution from exploring other paths that would lead to a failure in the same way.

{\em Craig interpolants}~\cite{Craig1957} allow deciding what information about a formula is relevant to a property. Assuming an implication $P\rightarrow Q$ holds in some logic, one can construct an interpolant $I$ such that $P\rightarrow I$ and $I\rightarrow Q$ are valid, and every non-logical symbol in $I$ occurs in both $P$ and $Q$. Interpolation is commonly used in program verification as follows: given a refutation proof for an unsatisfiable formula $P\wedge Q$, a {\em reverse interpolant} $I$ can be constructed such that $P\rightarrow I$ is valid and $I\wedge Q$ is unsatisfiable.
%In program verification, interpolants are typically devised as follows: given a refutation proof for an unsatisfiable formula $P\wedge Q$, an interpolant $I$ can be constructed such that $P\rightarrow I$ is valid and $I\wedge Q$ is unsatisfiable.

% aims at the falsification of safety properties of a program
Interpolation has largely been employed in model checking, predicate abstraction, predicate refinement, theorem proving, and other areas. For instance, interpolants provide a methodology to extend {\em bounded model checking} -- which aims at falsifying safety properties of a program for which the transition relation is unrolled up to a given bound -- to the unbounded case. In particular, since bounded proofs often contain the ingredients of unbounded proofs, interpolation can help construct an over-approximation of all reachable final states from the refutation proof for the bounded case, obtaining an over-approximation that is strong enough to prove absence of violations.
% given a refutation proof for the bounded case, 
% that is strong enough to provide guarantees on the absence of errors, but also loose enough to allow for an efficient computation.

\myparagraph{Subsumption with Interpolation} Interpolation can be used to tackle the path explosion problem when symbolically verifying programs marked (e.g., using assertions) with explicit error locations. As the exploration proceeds, the engine annotates each program location with conditions summarizing previous paths through it that have failed to reach an error location. Every time a branch is encountered, the executor checks whether the path conditions are subsumed by the previous explorations. In a best-case scenario, this approach can reduce the number of visited paths exponentially. %. Symbolic execution with interpolation has been proposed for software verification as an alternative to model checking-based techniques

%\cite{McMillan10} proposes an algorithm to annotate branches and statements with labels such that if they are implied by the current state, no error location can be reached from there.
\cite{McMillan10} proposes an annotation algorithm for branches and statements such that if their labels are implied by the current state, they cannot lead to an error location. Interpolation is used to construct weak labels that allow for an efficient computation of implication. \cite{YYG15} proposes a similar redundancy removal method called {\em postconditioned symbolic execution}, where  program locations are annotated with a postcondition, i.e., the {\em weakest precondition} summarizing path suffixes from previous explorations. The intuition here is that the weaker the interpolant is, the more likely it would enable path subsumption. Postconditions are constructed incrementally from fully explored paths and propagated backwards. When a branch is encountered, the corresponding postcondition is negated and added to the path constraints, which become unsatisfiable if the path is subsumed by previous explorations.

The soundness of path subsumption relies on the fact that an interpolant computed for a location captures the entirety of paths going through it. Thus, the path selection strategy plays a key role in enabling interpolant construction: for instance, DFS is very convenient as it allows exploring paths in full quickly, so that interpolants can be constructed and eventually propagated backwards; BFS instead hinders subsumption as interpolants may not available when checking for redundancy at branches as similar paths have not been explored in full yet. \cite{JMN13} proposes a novel strategy called {\em greedy confirmation} that decouples the path selection problem from the interpolant formation, allowing users to benefit from path subsumption when using heuristics other than DFS. Greedy confirmation distinguishes betweens nodes whose trees of paths have been explored in full or partially: for the latter, it performs limited traversal of additional paths to enable interpolant formation.

Interpolation has been proven to be useful for allowing the exploration of larger portions of a complex program within a given time budget. \cite{YYG15} claims that path redundancy is abundant and widespread in real-world applications. Typically, the overhead of interpolation - which can be performed within the SMT solver or in a dedicated engine - slows down the exploration in the early stages, then its benefits eventually start to pay off, allowing for a much faster exploration~\cite{JMN13}.

\myparagraph{Unbounded Loops} The presence of an unbounded loop in the code makes it harder to perform sound subsumption at program locations in it, as a very large number of paths can go through them. \cite{McMillan10} devises an iterative deepening strategy that unrolls loops until a fixed depth and tries to compute interpolants that are {\em loop invariant}, so that they can be used to prove the unreachability of error nodes in the unbounded case. This method however may not terminate for programs that require disjunctive loop invariants. \cite{JNS11} thus proposes a strategy to compute speculative invariants strong enough to make the symbolic execution of the loop converge quickly, but also loose enough to allow for path subsumption whenever possible. In a follow-up work~\cite{JMN12} loop invariants are discovered separately during the symbolic execution using a widening operator, and weakest preconditions for path subsumption are constructed such that they are entailed by the invariants.

We believe that the idea of using abstract interpretation in this setting -- originally suggested in~\cite{JSV09} -- deserves further investigation, as it can benefit from its many applications in other program verification techniques, and is amenable to an efficient implementation in mainstream symbolic executors, provided that the constructed invariants are accurate enough to capture the (un)rechability of error nodes.
%control-flow branches and program locations with labels, representing a condition under which no error locations can be reached. Labels are initially empty and constructed in a bottom-up fashion: once a path leading to an error has proven to be unfeasible,
%interpolation is used to compute a weaker formula that becomes the annotation for the last taken branch. As branches are explored,  
%has been fully explored and no error is found, interpolation is used to compute a weaker formula that 
%interpolation is used to compute conditions that are eventually propagated to their ancestors. 

\myparagraph{Subsumption with Abstraction} An approach not based on interpolation is taken in~\cite{APV09}, which describes a two-fold subsumption checking technique for symbolic states. A symbolic state is defined in terms of a symbolic heap and a set of constraints over scalar variables. The technique thus targets programs that manipulate not only scalar types, but also uninitialized or partially initialized data structures.  An algorithm for matching heap configurations through a graph traversal is presented, while an off-the-shelf solver is used to reason about subsumption for scalar data.

To cope with a possibly unbounded number of states, the work proposes abstraction to make the symbolic state space finite and thus subsumption effective.  Abstractions can summarize both the heap shape and the constraints on scalar data; examples are given for linked lists and arrays. Subsumption checking happens on under-approximate states, meaning that feasible behaviors could be missed. The authors employ the technique in a falsification scenario in combination with model checking, leaving to future work an application to verification based on symbolic execution only. 

%\myparagraph{Subsumption with Abstraction} A different approach not based on interpolation is taken in~\cite{APV09}, which proposes a {\em state matching} technique to compare symbolic states and to determine whether a state is subsumed by another one. The technique considers both heap shapes, by traversing the symbolic heap graphs in the symbolic states and trying to match their nodes, and state constraints due to numeric data stored in the symbolic states. State matching goes thus \mynote{how?} beyond bounded model checking and can handle un-initialized, or partially initialized, recursive data structures (such as linked lists or trees) as well as arrays.

%Even with subsumption, the number of symbolic states may still be unbounded. Hence, \cite{APV09} adds {\em abstractions} to limit the model checker's search space: for each explored state, the model checker computes and stores an abstract version of the state, as specified by suitable abstraction mappings, and subsumption checking is performed on the abstract states, effectively exploring an under-approximation of the feasible paths. 

\myparagraph{Path Partitioning} Dependence analyses for control and data flows expose casual relationships that one can use during the exploration to filter out paths unable to reveal additional program behavior. \cite{FLOWTEST-CAV09} partitions inputs for concolic execution in non-interfering blocks, symbolically exploring each block while others are kept fixed to concrete values. Interference of two inputs happens when they jointly affect one statement, or statements linked by control or data dependences. \cite{QNR13} focuses on outputs, placing two paths in the same partition if they have the same relevant slice with respect to the program output. A relevant slice is the transitive closure of dynamic data and control dependencies, and also of potential dependencies involving statements that affect the output by not getting executed. \cite{DGSE-TSE17} explores also faults irrelevant to the output by building relevant slices for individual statements, capturing how they are computed from symbolic inputs. A dependency analysis efficiently checks for equivalence of slices, deeming a path redundant when the slices for all its statement instances are collectively covered by previous paths.
% (which include also an additional kind named interactive)

% ---------------------------------------------------------------------------------------------------
\subsection{Under-constrained Symbolic Execution} 
\label{under-constrained}

% {\sc Check 'n' Crash}~\cite{CS-ICSE05}
A possible approach to avoid path explosion is to cut the code to be analyzed, say a function, out of its enclosing system and check it in isolation. Lazy initialization with user-specified preconditions (Section~\ref{ss:complex-objects}) follows this principle in order to automatically reconstruct complex  data structures. However, taking a code region out of an application may be quite difficult due to the entanglements with the surrounding environment~\cite{ED-ISSTA07}: errors detected in a function analyzed in isolation may be false positives, as the input may never assume certain values when the function is executed in the context of a full program. Some prior works, e.g., \cite{CS-ICSE05}, first analyze the code in isolation and then test the generated crashing inputs using concrete executions to filter out false positives.

%{\em Under-constrained symbolic execution}~\cite{ED-ISSTA07} is a twist on symbolic execution that allows for the analysis of a function in isolation by marking some symbolic inputs as {\em under-constrained}. Intuitively, a symbolic variable is under-constrained when in the analysis we do not account for constraints on its value that should have been collected along the path prefix from the program's entry point to the function to analyze. Under-constrained variables have the same semantics as classic symbolic variables except when used in an expression that can cause an error to occur. In particular, an error is reported only if all the solutions for the currently known constraints on the variable cause it to occur, i.e., the error is context-insensitive and thus a true positive. Otherwise, its negation is added to the path constraints and execution resumes as normal. This choice can be regarded as an attempt to reconstruct preconditions from the checks inserted in the code: any subsequent action violating an added negated constraint will be reported as an error.

% allows for
% expression that can cause an error to occur
{\em Under-constrained symbolic execution}~\cite{ED-ISSTA07} is a twist on symbolic execution that allows the analysis of a function in isolation by marking its symbolic inputs, as well as any global data that may affect its execution, as {\em under-constrained}. Intuitively, a symbolic variable is under-constrained when in the analysis we do not account for constraints on its value that should have been collected along the path prefix from the program's entry point to the function. In practice, a symbolic engine can automatically mark data as under-constrained without manual intervention by tracing memory accesses and identifying their location: e.g., a function's input can be detected when a memory read is performed on uninitialized data located on the stack. Under-constrained variables have the same semantics as classical fully constrained symbolic variables except when used in an expression that can yield an error. In particular, an error is reported only if all the solutions for the currently known constraints on the variable cause it to occur, i.e., the error is context-insensitive and thus a true positive. Otherwise, its negation is added to the path constraints and execution resumes as normal. This approach can be regarded as an attempt to reconstruct preconditions from the checks inserted in the code: any subsequent action violating an added negated constraint will be reported as an error. In order to keep this analysis correct, marks must be propagated between variables whenever any expression involves both under- and fully constrained values. For instance, a comparison of the form {\tt a > b}, where {\tt a} is under-constrained and {\tt b} is not, forces the engine to propagate the mark from {\tt a} to {\tt b}, similarly as in taint analysis when handling tainted values. Marks are typically tracked by the symbolic engine using a shadow memory.

%{\em Under-constrained symbolic execution}~\cite{ED-ISSTA07} is a twist on symbolic execution that allows for the analysis of a function in isolation by marking symbolic inputs for which preconditions are missing as {\em under-constrained}. Intuitively, missing preconditions are the constraints on the variable yielded along the path prefix from the program's entry point to the function. Under-constrained variables have the same semantics as classic symbolic variables except when used in an expression that can cause an error to occur. In this case, an error is reported only if all the solutions for the currently known constraints on the variable cause it to occur, i.e., the error is context-insensitive and a true positive. Otherwise, its negation is added to the path constraints and execution resumes as normal. This choice can be regarded as an attempt to reconstruct preconditions from the checks inserted in the code. Any subsequent action violating an added negated constraint will be reported as an error.

Although this technique is not sound as it may miss errors, it can still scale to find interesting bugs in larger programs. Also, the application of under-constrained symbolic execution is not limited to functions only: for instance, if a code region (e.g., a loop) may be troublesome for the symbolic executor, it can be skipped by marking the locations it affects as under-constrained. Since in general it is not easy to understand which data could be affected by the execution of some skipped code, manual annotation may be needed in order to keep the analysis correct.

% ---------------------------------------------------------------------------------------------------
\subsection{Exploiting Preconditions and Input Features}%\mynote{[D] Entirely rewritten}
\label{precontioned-symbolic-execution}

% input state space
Another way to reduce the path explosion is to leverage knowledge of some input properties. {\sc AEG}~\cite{AEG-NDSS11} proposes {\em preconditioned symbolic execution} to reduce the number of explored states by directing the exploration to a subset of the input space that satisfies a precondition predicate. The rationale is to focus on inputs that may lead to certain behaviors of the program (e.g., narrowing down the exploration to inputs of maximum size to reveal potential buffer overflows). Preconditioned symbolic execution trades soundness for performance: well-designed preconditions should  be neither too specific (they would miss interesting paths) nor too generic (they would compromise the speedups resulting from the space state reduction). Instead of starting from an empty path constraints set, the approach adds the preconditions to the initial $\pi$ so that the rest of the exploration will skip branches that do not satisfy them. While adding more constraints to $\pi$ at initialization time is likely to increase the burden on the solver, required to perform a larger number of checks at each branch, this may be largely outweighted by the performance gains due to the smaller state space.

Common types of preconditions considered in symbolic execution are: {\em known-length} (i.e., the size of a buffer is known), {\em known-prefix} (i.e., a buffer has a known prefix), and {\em fully known} (i.e., the content of a buffer is fully concrete). These preconditions are rather natural when dealing with code that operates over inputs with a well-known or predefined structure, such as string parsers or packet processing tools. 

\begin{shaded}\noindent{\bf\small Example.} Consider the following simplified packet header processing code: {\tt pkt} points to the input buffer, while {\tt header} to the fixed expected content. If no
\begin{wrapfigure}{r}{0.4\textwidth}
  \vspace{-4.2mm}
  \begin{lstlisting}[basicstyle=\ttfamily\small]
 start: get_input(&pkt);
 for(k = 0; k < 128; k++)
   if (pkt[k] != header[k])
     goto start;
 parse_payload(&pkt)
\end{lstlisting}
\vspace{-5.8mm}
\end{wrapfigure}
precondition is considered, then this code can generate an exponential number of paths since any mismatch forces a new call to {\tt get\_input}. On the other hand, if a {\em known prefix} precondition is set on the input, then only a single path is generated when exploring the loop. The engine can thus focus its exploration on {\tt parse\_payload()}.
\end{shaded}

%By using static or dynamic analysis techniques, it may be possible to derive properties over a loop that can be exploited by the symbolic engine to significantly prune branching paths. For instance, knowledge of the exact number of loop iterations - or at least an upper bound on it - can significantly help the engine.  

%\noindent
Of a different flavor is the work by~\cite{SPM-ISSTA09}, which presents a technique, called {\em loop-extended symbolic execution}, that is able to effectively explore a loop whenever a grammar describing the input program is available. Relating the number of iterations with features of the program input can profitably guide the exploration of the program states generated by a loop, reducing the path explosion problem.
%Based on the observation that loop executions may strictly depend on input features, \cite{SPM-ISSTA09} presents a technique, called {\em loop-extended symbolic execution}, which is able to effectively explore a loop whenever a grammar describing the input program is available. Relating the number of iterations with features of the program input can profitably guide the exploration of the program states generated by a loop.

% ---------------------------------------------------------------------------------------------------

%\subsection{Dynamic symbolic execution}
%Dynamic symbolic execution refers to a body of techniques that exploit execution with concrete values to explore [...].

% ---------------------------------------------------------------------------------------------------

\subsection{State Merging}
State merging is a powerful technique that fuses different paths into a single state. A merged state is described by a formula that represents the disjunction of the formulas that would have described the individual states if they were kept separate. Differently from other static program analysis techniques such as abstract interpretation, merging in symbolic execution does not lead to over-approximation.

%Several static program analysis techniques such as abstract interpretation merge states corresponding to different paths into a state that over-approximates them. The notion of state merging was also explored in the context of symbolic execution
%In a precise symbolic execution, however, merging is not allowed \mynote{over--approxim?} to introduce any approximation or abstraction, and therefore can only change formulas to have them characterize sets of execution paths. 
%In other words, a merged state will be described by a formula that represents the disjunction of the formulas that would have described the individual states if they were kept separate.

\begin{shaded}\noindent{\bf\small Example.} Consider function {\tt foo} shown below and its symbolic execution tree shown in Figure~\ref{fig:example-state-merging}a. Initially (execution state $A$) the path constraints are {\em true} and input arguments {\tt x} and {\tt y} are associated with symbolic values $\alpha_x$ and $\alpha_y$, respectively.
\begin{wrapfigure}{l}{0.4\textwidth}
  \vspace{-4.2mm}
  \begin{lstlisting}[basicstyle=\ttfamily\small]
 1. void foo(int x, int y) {
 2.  if (x < 5)
 3.   y = y * 2;
 4.  else
 5.   y = y * 3;
 6.  return y;
 7. }
\end{lstlisting}
\vspace{-5.8mm}
\end{wrapfigure}
After forking due to the conditional branch at line 2, a different statement is evaluated and different assumptions are made on symbol $\alpha_x$ (states $B$ and $C$, respectively). When the {\tt return} at line 6 is eventually reached on all branches, the symbolic execution tree gets populated with two additional states, $D$ and $E$. In order to reduce the number of active states, the symbolic engine can perform state merging. For instance, Figure~\ref{fig:example-state-merging}b shows the symbolic execution DAG for the same piece of code when a state merging operation is performed before evaluating the {\tt return} at line 6: $D'$ is a merged state that fully captures the former execution states $D$ and $E$ using the $ite$ expression $ite(\alpha_x<5, 2*\alpha_y, 3*\alpha_y)$ (Section~\ref{ss:fully-symbolic-memory}). 
%Indeed, the $ite(\texttt{c}, \texttt{t}, \texttt{f})$ expression introduced in the symbolic store $\sigma$ is a short term for an {\tt if-then-else} expression and means that if the condition {\tt c} is verified then {\tt t} holds, otherwise {\tt f} must be assumed as true. Nonetheless, $ite$ expressions are often just syntactic sugar for disjunctive formulas and are commonly supported by most prominent constraint solvers. For instance, in the context of propositional logic the $ite(\texttt{c}, \texttt{t}, \texttt{f})$  expression could be replaced with the formula $(\texttt{c} \wedge \texttt{t}) \vee (\neg\texttt{c} \wedge \texttt{f})$ . However, since the symbolic store in our model should return an integer value for variable $y$ rather than a boolean value, following the idea presented in~\cite{KP-PP05}, the $ite$ expression could be translated into the expression $((\alpha_x < 5) * (2 * \alpha_y)) + (\neg(\alpha_x < 5) * (3 * \alpha_y))$ that evaluates\footnote{We are assuming that the result of a comparison maps to integer values 0 or 1.} to the actual value of {\tt y} based on the branch condition at line 2. 
%Indeed, the condition $(\alpha_x < 5)$ could be either true or false, yielding to only one of two possible values of $y$. 
Note that the path constraints of the execution states $D$ and $E$ can be merged into the disjunction formula $\alpha_x < 5 \vee \alpha_x \geq 5$ and then simplified to $true$ in $D'$.
\end{shaded}

\begin{figure}[t]
  %\vspace{-3mm}
  \centering
  \begin{subfigure}{.5\textwidth}
    \centering
    \hspace{-5mm}
    \includegraphics[width=1.05\columnwidth]{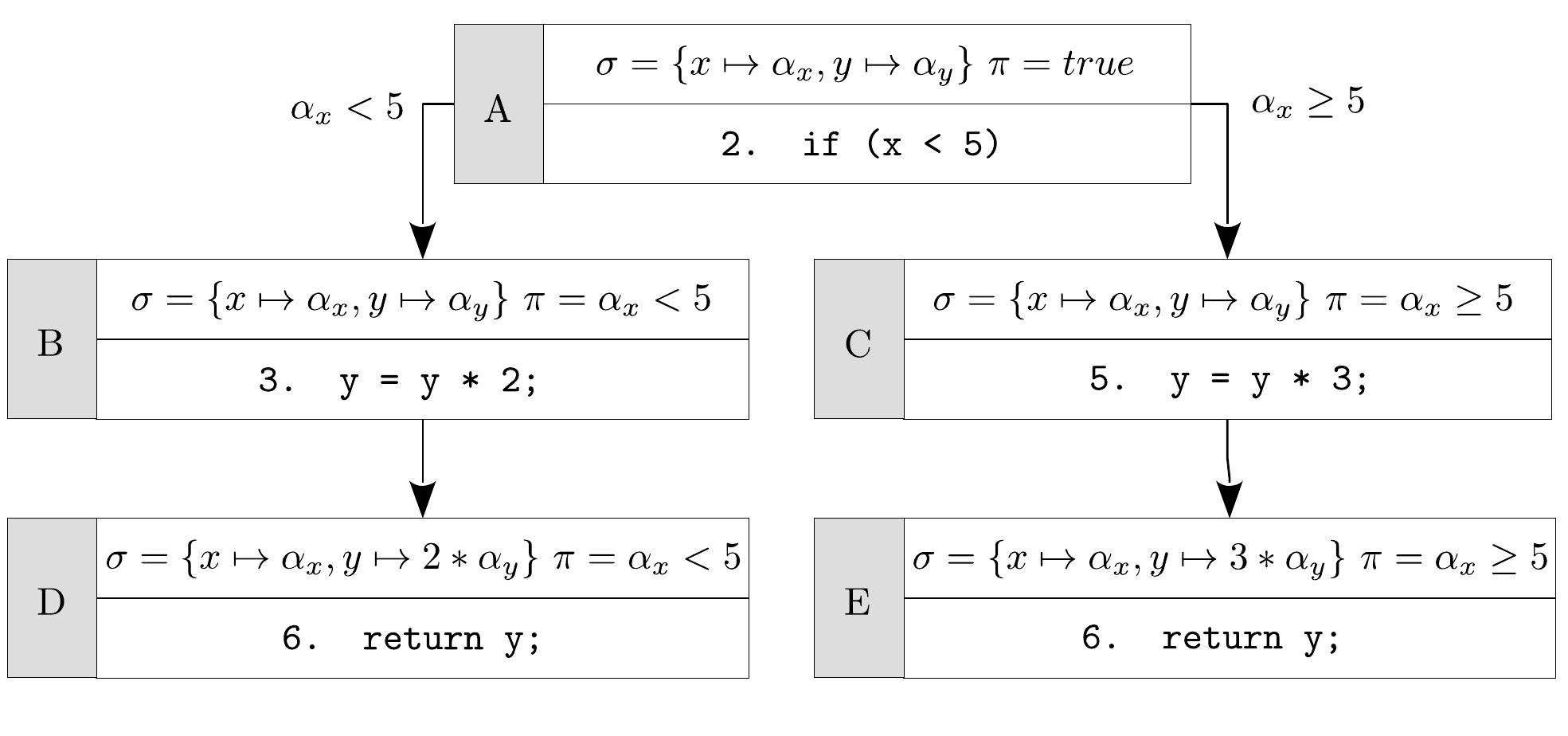} 
    %\label{fig:sub1}
    \vspace{-6.5mm}
    \caption{}
  \end{subfigure}%
  \begin{subfigure}{.5\textwidth}
    \centering
    \includegraphics[width=1.05\columnwidth]{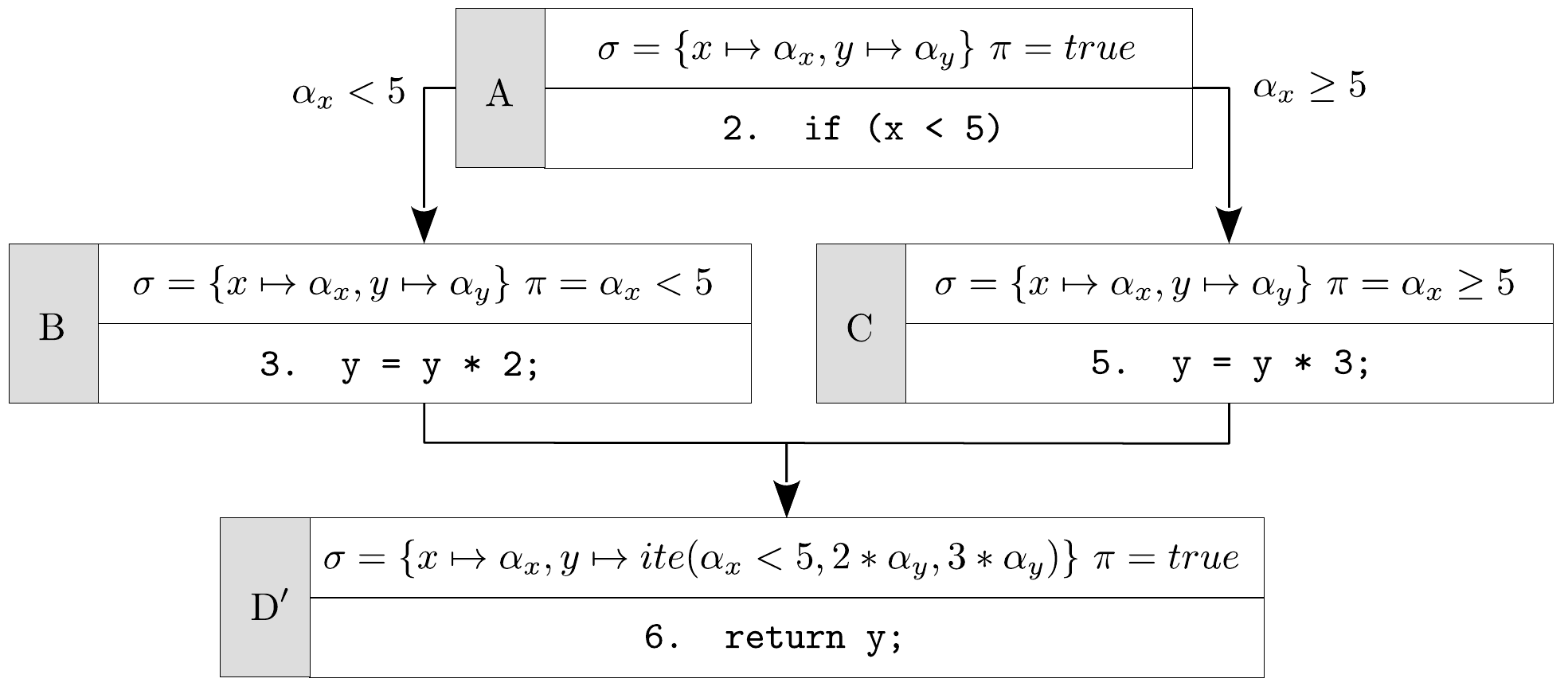} 
    %\label{fig:sub2}
    \vspace{-4mm}
    \caption{}
  \end{subfigure}
  \vspace{-3mm}
  \caption{Symbolic execution of function \texttt{foo}: (a) without and (b) with state merging.}
  \label{fig:example-state-merging}
\end{figure}

%\vspace{-4pt} % TODO
% $(stmt_1,~\sigma_1,~\pi_1)$ and $(stmt_1,~\sigma_2,~\pi_2)$, the merged state can be constructed as $(stmt,~\sigma',~\pi_1 \vee \pi_2)$ where $stmt = stmt_1 = stmt_2$
\myparagraphnoperiod{Tradeoffs: to Merge or Not to Merge?} In principle, it may be profitable to apply state merging whenever two symbolic states about to evaluate the same statement are very similar (i.e., differ only for few elements) in their symbolic stores. Given two states $(stmt,~\sigma_1,~\pi_1)$ and $(stmt,~\sigma_2,~\pi_2)$, the merged state can be constructed as $(stmt,~\sigma',~\pi_1 \vee \pi_2)$, where $\sigma'$ is the merged symbolic store between $\sigma_1$ and $\sigma_2$ built with {\em ite} expressions accounting for the differences in storage, while $\pi_1 \vee \pi_2$ is the union of the path constraints from the two merged states. Control-flow structures such as if-else statements (as in the previous example) or simple loops often yield rather similar successor states that represent very good candidates for state merging.

Early works~\cite{G-POPL07,HSS-RV09} have shown that merging techniques effectively decrease the number of paths to explore, but also put a burden on constraints solvers, which can be hampered by disjunctions. Merging can also introduce new symbolic expressions in the code, e.g., when merging different concrete values from a conditional assignment into a symbolic expression over the condition. \cite{KKB-PLDI12} provides an excellent discussion of the design space of state merging techniques. At the one end of the spectrum, complete separation of paths used in search-based symbolic execution (Section~\ref{ss:heuristics}) performs no merge. At the other end, static state merging combines states at control-flow join points, essentially representing a whole program with a single formula. Static state merging is used in whole-program verification condition generators\iffullver{, e.g.,} ~\cite{SATURN-POPL05,CALYSTO-ICSE08}), \iffullver{which typically trade precision for scalability by, for instance, unrolling loops only once.}{which usually trade precision for scalability e.g., by unrolling loops only once.}

%At the other end, static state merging combines states at control-flow join points after all subpaths leading to a join point have been explored.

%\mynote{Function summaries}

\myparagraph{Merging Heuristics} Intermediate merging solutions adopt heuristics to identify state merges that can speed the exploration process up. Indeed, generating larger symbolic expressions and possibly extra solvers invocations can outweigh the benefit of having fewer states, leading to poorer overall performance~\cite{HSS-RV09,KKB-PLDI12}. {\em Query count estimation}~\cite{KKB-PLDI12} relies on a simple static analysis to identify how often each variable is used in branch conditions past any given point in the CFG. The estimate is used as a proxy for the number of solver queries that a given variable is likely to be part of. Two states make a good candidate for merging when their differing variables are expected to appear infrequently in later queries. {\em Veritesting}~\cite{VERITESTING-ICSE14} implements a form of merging heuristic based on a distinction between easy and hard statements, where the latter involve indirect jumps, system calls, and other operations for which precise static analyses are difficult to achieve. Static merging is performed on sequences of easy statements, %which are represented as a single path 
whose effects are captured using $ite$ expressions, 
while per-path symbolic exploration is done whenever a hard-to-analyze statement is encountered. 

%identifies sequences of statements that do not contain system calls, indirect jumps, and other statements that are difficult to reason about statically, and represents them with a single formula as in static state merging. The approach is alternated with a per-path basis symbolic exploration every time a hard-to-analyze statement is encountered. 

\myparagraph{Dynamic State Merging} In order to maximize merging opportunities, a symbolic engine should traverse a CFG so that a combined state for a program point can be computed from its predecessors, e.g., if the graph \mynote{check} is acyclic, by following a topological ordering. However, this would prevent search exploration strategies that prioritize ``interesting'' states. \cite{KKB-PLDI12} introduces {\em dynamic state merging} which works regardless of the exploration order imposed by the search strategy.
Suppose the symbolic engine maintains a worklist of states and a bounded history of their predecessors. When the engine has to pick the next state to explore, it first checks whether there are two states $s_1$ and $s_2$ from the worklist such that they do not match for merging, but $s_1$ and a predecessor of $s_2$ do. If the expected similarity between $s_2$ and a successor of $s_1$ is also high, the algorithm attempts a merge by advancing the execution of $s_1$ for a fixed number of steps. This captures the idea that if two states are similar, then also their respective successors are likely to become similar in a few steps. If the merge fails, the algorithm lets the search heuristic pick the next state to explore.

%This is useful, for instance, for unbounded loops for which search-based symbolic execution engines would employ search strategies that prioritize exploring new code over unrolling, while static state merging would require a depth-first exploration and thus fully unroll the possibly infinitely many iterations of the loop.

% ---------------------------------------------------------------------------------------------------
\subsection{Leveraging Program Analysis and Optimization Techniques}
\label{ss:program-analysis}

A deeper understanding of a program's behavior can help a symbolic engine optimize its analysis and focus on promising states, e.g., by pruning uninteresting parts of the computation tree. Several classical program analysis techniques have been explored in the symbolic execution literature. We now briefly discuss some prominent examples.

% \mytempedit{In this section we explore connections with other program analysis and verification techniques.}

%is a method that, starting from a subset of a program's behavior, extracts from the program the minimal sequence of instructions that faithfully represents that behavior~\cite{Weiser84}.
\myparagraph{Program Slicing} This analysis, starting from a subset of a program's behavior, extracts from the program the minimal sequence of instructions that faithfully represents that behavior~\cite{Weiser84}. This information can help a symbolic engine in several ways: for instance, \cite{FIRMALICE-NDSS15} exploits backward program slicing to restrict symbolic exploration toward a specific target program point.

 %given a program slice related to a target program point, symbolic exploration can be restricted to paths contained in the program slice. \iffullver{We discuss an example of use in Section~\ref{ss:auth-bypass}.}{}

\myparagraph{Taint Analysis} This technique~\cite{SAB-SP10} attempts to check which variables of a program may hold values derived from potentially dangerous external sources such as user input. The analysis can be performed both statically and dynamically, with the latter yielding more accurate results. In the context of symbolic execution, taint analysis can help an engine detect which paths depend on tainted values. For instance,~\cite{MAYHEM-SP12} focuses its analysis on paths where a jump instruction is tainted and uses symbolic execution to generate an exploit.
% skip execution paths that do not depend upon tainted values, effectively reducing the exploration state space.}
%In the context of symbolic execution, taint analysis can help an engine skip execution paths that do not depend upon tainted values, effectively reducing the exploration state space~\cite{SAB-SP10}.

\myparagraph{Fuzzing} This software testing approach randomly mutates user-provided test inputs to cause crashes or assertion failures, possibly finding potential memory leaks. Fuzzing can be augmented with symbolic execution to collect constraints for an input and negate them to generate new inputs. On the other hand, a symbolic executor can be augmented with fuzzing to reach deeper states in the exploration more quickly and efficiently. Two notable embodiments of this idea are represented by {\em hybrid concolic testing}~\cite{RK-ICSE07} and Driller~\cite{DRILLER-NDSS16}.
%\iffullver{: we present two embodiments of this approach in Section~\ref{ss:bug-detection}.} 

%{.\mynote{CD: consider citing fuzzers formerly in the apps section}}
  %\mynote{C: crossref: dynamic test generation}
  %\footnote{\cite{DRILLER-NDSS16} classifies offline symbolic executors such as {\sc DART} and {\sc SAGE} as {\em whitebox fuzzers}.}

\myparagraph{Branch Predication} This is a strategy for mitigating misprediction penalties in pipelined executions by avoiding jumps over very small sections of code: for instance, control-flow forking constructs such as the C ternary operator can be replaced with a predicated {\tt select} instruction. \cite{CCK-EUROSYS11} reports an exponential decrease in the number of paths to explore from the adoption of this strategy when cross-checking two implementations of a program using symbolic execution. % [D] alternative formulation: ~\cite{CCK-EUROSYS11} cross-checks two implementations of a program using symbolic execution and reports an exponential decrease in the number of paths to explore from the adoption of a simple form of this strategy; 
%  \item {\em source code analysis}: \mynote{TODO/drop?} extraction of input properties (e.g., size or contents of an array);

\myparagraph{Type Checking} Symbolic analysis can be effectively mixed with typed checking~\cite{KCF-PLDI10}:  for instance, type checking can determine the return type of a function that is difficult to analyze symbolically: such information can then potentially be used by the executor to prune certain paths\footnote{The work also discusses how a symbolic analysis can help type checking, e.g, by providing context-sensitive properties over a variable that would rule out certain type errors, improving the precision of the type checker.}.

%\mytempedit{\myparagraph{State matching} This approach determines whether an abstract state is subsumed by another}, and can be used to analyze an under-approximation of a program's behavior. For instance, \cite{APV-SPIN06,VPP-ISSTA06} explore different heap shapes in the context of test generation for data structures, using subsumption checking to determine whether a symbolic state is being revisited. % ~\cite{XGM-ISSTA08} implements a different form of under-approximation: it looks for buffer overflows by having only a prefix of a buffer handled symbolically, and a symbolic length that may exceed the lenght of the prefix (any byte beyond the prefix is filled with concrete random data) - [D] I'm dropping this for now as we would have to call the item 'under-approximation' and change the first part of the text

\myparagraph{Program Differencing} Dependence analyses can identify branches and data flows affected by code edits. {\em Directed incremental symbolic execution}~\cite{DISE-TOSEM14} statically identifies CFG nodes affected by changes, and uses such information to drive the exploration to only those paths that exercise uncovered sequences of affected nodes.
% pruning paths that would result in path conditions unaffected by changes.}

\myparagraph{Compiler Optimizations}
\cite{Cadar-FSE15} argues that program optimization techniques should be a first-class ingredient of practical implementations of symbolic execution, alongside widely accepted solutions such as search heuristics, state merging, and constraint solving optimizations. In fact, program transformations can affect both the complexity of the constraints generated during path exploration and the exploration itself. For instance, precomputing the results of a function using a lookup table leads to a larger number of constraints in the path conditions due to memory accesses, while applying strength reduction for multiplication may result in a chain of addition operations that is more expensive for a constraint solver. Also, the way high-level {\tt switch} statements are compiled can significantly affect the performance of path exploration, while resorting to conditional instructions such as {\tt select} in LLVM or {\tt setcc} and {\tt cmov} in x86 can avoid expensive state forking by yielding simple {\em ite} expressions instead.
% for a multiplication by a constant

% D-THESIS14 superseded by DOZ-ISSRE15
%and none of them has tackled the analysis of binary programs.
While the effects of a compiler optimization can usually be predicted on the number or size of the instructions executed at run time, a similar reduction is not obvious in symbolic execution~\cite{DOZ-ISSRE15}, mostly because the constraint solver is typically used as a black-box. To the best of our knowledge, only a few works have attempted to analyze the impact of compiler optimizations on constraint generation and path exploration~\cite{WKC-HOTOS13,DOZ-ISSRE15}, leaving interesting open questions. \mynote{TODO: Move to 3.1?}Of a different flavor is the work presented in~\cite{PMZ-ISSTA17}, which explores transformations such as dynamic constant folding and optimized constraint encoding to speed up memory operations in symbolic executors based on theories of arrays (Section~\ref{ss:fully-symbolic-memory}).

%% file: constraints.tex
% !TEX root = main.tex

\section{Constraint solving}
\label{se:constraint-solving}

Constraint satisfaction problems arise in many domains, including analysis, testing, and verification of software programs. Constraint solvers are decision procedures for problems expressed in logical formulas: for instance, the boolean satisfiability problem (also known as SAT) aims at determining whether there exists an interpretation of the symbols of a formula that makes it true. Although SAT is a well-known NP-complete problem, recent advances have moved the boundaries for what is intractable when it comes to practical applications~\cite{SMT-CACM11}. 

% linear arithmetic inequalities
Observe that some problems are more naturally described with languages that are more expressive than the one of boolean formulas with logical connectives. For this reason, satisfiability modulo theories (SMT) generalize the SAT problem with supporting theories to capture formulas involving, for instance, linear arithmetic and operations over \iffullver{arrays (see, e.g., Section~\ref{ss:fully-symbolic-memory}).}{arrays.} SMT solvers map the atoms in an SMT formula to fresh boolean variables: a SAT decision procedure checks the rewritten formula for satisfiability, and a theory solver checks the model generated by the SAT procedure.

%\mytempedit{In particular, SMT-compliant theory solvers are required to be able to: (i) work incrementally when checking for consistency as novel constraints are added, (ii) support backtracking, i.e., constraint removal, and (iii) provide explanations for inconsistent constraints~\cite{Abraham15}.}

SMT solvers show several distinctive strengths. Their core algorithms are generic, and can handle complex combinations of many individual constraints. They can work incrementally and backtrack as constraints are added or removed, and provide explanations for inconsistencies. Theories can be added and combined in arbitrary ways, e.g., to reason about arrays of strings. Decision procedures do not need to be carried out in isolation: often, they are profitably combined to reduce the amount of time spent in heavier procedures, e.g., by solving linear parts first in a non-linear arithmetic formula. Incomplete procedures are valuable too: complete but expensive procedures get called only when conclusive answers could not be produced. All these factors allows SMT solvers to tackle large problems that no single procedure can solve in isolation\footnote{We refer the interested reader to~\cite{BKM14} for an exhaustive introduction to SMT solving, and to~\cite{SC2} for a discussion of its distinctive strengths.}.
% SHORTER VERSION
% }%\footnote{\cite{BKM14,SC2} provide interesting discussions of the strengths of SMT solvers.}.}

%\mytempedit{SMT solvers show a number of distinctive strengths. They can work incrementally as constraints are added to formulas, backtrack for constraint removal, and provide explanations for inconsistent constraints. Their core algorithms are generic and can handle complex combinations of many individual constraints. Theories can be added and, more importantly, combined in arbitrary ways, e.g., to reason about arrays of strings. Decision procedures are not required to be carried out in isolation: often, they can profitably be combined to reduce the amount of time spent in heavier procedures, e.g., by solving linear problem parts first for a non-linear arithmetic formula. Incomplete procedures are valuable too: complete but expensive procedures get called only when conclusive answers could not be produced. The combination of these factors allows SMT solvers to tackle large problems that no single procedure can solve in isolation\footnote{We refer the interested reader to~\cite{BKM14} for an exhaustive introduction to SMT solving, and to~\cite{SC2} for a discussion of its distinctive strengths.}.}

% STP~\cite{STP-CAV07,STP-TR07} solver
% {\sc MineSweeper}~\cite{MineSweeper-BOTNET08}, and {\sc AEG}~\cite{AEG-NDSS11}
In a symbolic executor, constraint solving plays a crucial role in checking the feasibility of a path, generating assignments to symbolic variables, and verifying assertions.
Over the years, different solvers have been employed by symbolic executors, depending on the supported theories and the relative performance at the time. For instance, the STP~\cite{STP-CAV07} solver has been employed in, e.g., {\sc EXE}~\cite{EXE-CCS06}, {\sc KLEE}~\cite{KLEE-OSDI08}, and {\sc AEG}~\cite{AEG-NDSS11}, which all leverage its support for bit-vector and array theories. Other executors such as {\sc Java PathFinder}~\cite{PATHFINDER-ASE10} have complemented SMT solving with additional decision procedures (e.g., libraries for constraint programming~\cite{CHOCO}) and heuristics to handle complex non-linear mathematical constraints~\cite{CORAL-NFM11}.

Recently, Z3~\cite{Z3-TACS08} has emerged as leading solution for SMT solving. Developed at Microsoft Research, Z3 offers cutting-edge performance and supports a large number of theories, including bit-vectors, arrays, quantifiers, uninterpreted functions, linear integer and real arithmetic, and non-linear arithmetic. 
%
%Effective support for strings has been recently offered by Z3-str~\cite{ZZG-FSE13}, an extension of Z3 that makes it possible to treat string as a primitive type, allowing the solver to reason on common string operations such as concatenation, substring, and replacement.
Its Z3-str~\cite{ZZG-FSE13} extension makes it possible to treat also strings as a primitive type, allowing the solver to reason on common string operations such as concatenation, substring, and replacement.
Z3 is employed in most recently appeared symbolic executors such as {\sc Mayhem}~\cite{MAYHEM-SP12}, {\sc SAGE}~\cite{SAGE-QUEUE12}, and {\sc Angr}~\cite{ANGR-SSP16}. Due to the extensive number of supported theories in Z3, such executors typically do not to employ additional decision procedures.

%The two most popular solvers used in symbolic executors are STP and Z3. STP~\cite{STP-CAV07,STP-TR07} is an SMT solver with bitvector and array theories initially developed at Stanford and employed in, e.g., {\sc EXE}~\cite{EXE-CCS06}, {\sc KLEE}~\cite{KLEE-OSDI08}, {\sc MineSweeper}~\cite{MineSweeper-BOTNET08}, and {\sc AEG}~\cite{AEG-NDSS11}. Z3~\cite{Z3-TACS08} is an SMT solver developed at Microsoft with support for nonlinear arithmetic, bitvector, and array theories, and is used in, e.g., {\sc Mayhem}~\cite{MAYHEM-SP12}, {\sc SAGE}~\cite{SAGE-QUEUE12}, and {\sc Angr}~\cite{ANGR-SSP16}. CVC3~\cite{CVC3-CAV07} is another SMT solver that supports theories for linear arithmetic, bitvectors, arrays, and quantifiers, and is employed in {\sc Java PathFinder}~\cite{PATHFINDER-ASE10} along with CHOCO~\cite{CHOCO} for integer/real constraints and CORAL~\cite{CORAL-NFM11} for complex mathematical constraints. Modern symbolic executors can typically choose between different underlying solvers through a common API, and also resort to a native interface to a specific solver for better performance.

%only for efficiency reasons.

%For instance, many solvers have the development of ~\cite{PATHFINDER-ASE10} can use a large number of SMT solvers, including Yices, 
%~\cite{YICES-CAV06} is an incremental solver with support for rational and integer linear arithmetic, bitvectors, and arrays, and was originally used in 
%In Table~\ref{tab:solvers} we report a number of constraint solving tools used in popular symbolic execution engines.

% feasibility or applicability? TODO
However, despite the significant advances observed over the past few years -- which also made symbolic execution practical in the first place~\cite{CS-CACM13} -- constraint solving remains one of the main obstacles to the scalability of symbolic execution engines, and also hinders its feasibility in the face of constraints that involve expensive theories (e.g., non-linear arithmetic) or opaque library calls.

%\subsection{Optimization Techniques}
%\label{ss:constraint-opt}

% handling or skipping over
In the remainder of this section, we address different techniques to extend the range of programs \iffullver{that can be handled by}{amenable to} symbolic execution and to optimize the performance of constraint solving. Prominent approaches consist in: (i) reducing the size and complexity of the constraints to check, (ii) unburdening the solver by, e.g., resorting to constraint solution caching, deferring of \iffullver{constraint solver queries}{solver queries}, or concretization, and (iii) augmenting symbolic execution to handle constraints problematic for decision procedures.

%We conclude by pointing out potential directions to improve support for non-linear arithmetic}.

%\mytempedit{and (iii) augmenting symbolic execution with techniques aimed at handling constraints that are problematic for the underlying decision procedure. We conclude the section by pointing out potential research directions to improve support for non-linear arithmetic}.

%: (i) {\em constraint reduction} techniques aim at simplifying constraints fed to a solver by rewriting them into a shorter form: (ii) techniques for {\em reuse of constraint solutions} explore the space-time trade-off of retrieving previously computed query results rather than repeating expensive satisfiability checks.

\myparagraph{Constraint Reduction} 
A common optimization approach followed by both solvers and symbolic executors is to reduce constraints into simpler forms. For example, the {\em expression rewriting} optimization can apply classical techniques from optimizing compilers such as constant folding, strength reduction, and simplification of linear expressions (see, e.g., {\sc KLEE}~\cite{KLEE-OSDI08}).

{\sc EXE}~\cite{EXE-CCS06} introduces a {\em constraint independence} optimization that exploits the fact that a set of constraints can frequently be divided into multiple independent subsets of constraints. This optimization interacts well with query result caching strategies, and offers an additional advantage when an engine asks the solver about the satisfiability of a specific constraint, as it removes irrelevant constraints from the query. In fact, independent branches, which tend to be frequent in real programs, could lead to unnecessary constraints that would get quickly accumulated.

Another fact that can be exploited by reduction techniques is that the natural structure of programs can lead to the introduction of more specific constraints for some variables as the execution proceeds. Since path conditions are generated by conjoining new terms to an existing sequence, it might become possible to rewrite and optimize existing constraints. For instance, adding an equality constraint of the form $x:=5$ enables not only the simplification to true of other constraints over the value of the variable (e.g., $x>0$), but also the substitution of the symbol $x$ with the associated concrete value in the other subsequent constraints involving it. The latter optimization is also known as {\em implied value concretization} and, for instance, it is employed by {\sc KLEE}~\cite{KLEE-OSDI08}.

In a similar spirit, {\sc \stwoe}~\cite{CKC-TOCS12} introduces a bitfield-theory expression simplifier to replace with concrete values parts of a symbolic variable that bit operations mask away. For instance, for any 8-bit symbolic value $v$, the most significant bit in the value of expression $v\,|\,10000000_2$ is always 1. The simplifier can propagate information across the tree representation of an expression, and if each bit in its value can be determined, the expression is replaced with the corresponding constant.
 
%path conditions in a symbolic executor are typically generated by conjoining a new term to an existing (and possibly satisfiable) sequence of constraints. As the exploration proceeds, the natural structure of programs means that constraints might become more specific for some variables, and constraints can be rewritten accordingly. 

%\subsubsection{Reuse of Constraint Solutions}
%\label{ss:constraint-reuse}

%\subsection{Unburdening the Constraint Solver} 
%\label{ss:solver-unburdening}

\myparagraph{Reuse of Constraint Solutions} 
The idea of reusing previously computed results to speed up constraint solving can be particularly effective in the setting of a symbolic executor, especially when combined with other techniques such as constraint independence optimization. Most reuse approaches for constraint solving are currently based on semantic or syntactic equivalence of the constraints.

{\sc EXE}~\cite{EXE-CCS06} caches the results of constraint solutions and satisfiability queries in order to reduce as much as possible the need for calling the solver. A cache is handled by a server process that can receive queries from multiple parallel instances of the execution engine, each exploring a different program state.

{\sc KLEE}~\cite{KLEE-OSDI08} implements an incremental optimization strategy called {\em counterexample caching}. Using a cache, constraint sets are mapped to concrete variable assignments, or to a special null value when a constraint set is unsatisfiable. When an unsatisfiable set in the cache is a subset for a given constraint set $S$, $S$ is deemed unsatisfiable as well. Conversely, when the cache contains a solution for a superset of $S$, the solution trivially satisfies $S$ too. Finally, when the cache contains a solution for one or more subsets of $S$, the algorithm tries substituting in all the solutions to check whether a satisfying solution for $S$ can be found.

{\em Memoized symbolic execution}~\cite{MEMO-ISSTA12} is motivated by the observation that symbolic execution often results in re-running largely similar sub-problems, e.g., finding a bug, fixing it, and then testing the program again to check if the fix was effective. The taken choices during path exploration are compactly encoded in a prefix tree, opening up the possibility to reuse previously computed results in successive runs.
%  in a trie-based data structure

The Green framework~\cite{GREEN-FSE12} explores constraint solution reuse across runs of not only the same program, but also similar programs, different programs, and different analyses. Constraints are distilled into their essential parts through a {\em slicing} transformation and represented in a canonical form to achieve good reuse, even within a single analysis run. \cite{JGY-ISSTA15} presents an extension to the framework that exploits logical implication relations between constraints to support constraint reuse and faster execution times.

%\subsection{Other Optimizations in Symbolic Executors}
%\subsection{Reducing the Symbolic Executor's Pressure on Constraint Solvers}
%\label{ss:reducing-constraint-solver-pressure}

%In this section we present a number of other optimizations that become possible in the setting of a symbolic executor to reduce the time spent in the constraint solver.

\myparagraph{Lazy Constraints}
\cite{UCKLEE-USEC15} adopts a timeout approach for constraint solver queries. In their initial experiments, the authors traced most timeouts to symbolic division and remainder operations, with the worst cases occurring when an unsigned remainder operation had a symbolic value in the denominator.
They thus implemented a solution that works as follow: when the executor encounters a branch statement involving an expensive symbolic operation, it will take both the true and false branches and add a {\em lazy} constraint on the result of the expensive operation to the path conditions. When the exploration reaches a state that satisfies some goal (e.g., an error is found), the algorithm will check for the feasibility of the path, and suppress it if deemed unreachable in a real execution.

Compared to the {\em eager} approach of checking the feasibility of a branch as encountered (Section~\ref{ss:unrealizable-paths}), a lazy strategy may lead to a larger number of active states, and in turn to more solver queries. However, the authors report that the delayed queries are in many cases more efficient than their eager counterparts: the path constraints added after a lazy constraint can in fact narrow down the solution space for the solver.

\begin{figure}[t]
  \begin{center}
  \begin{subfigure}{.43\textwidth}
    %\vspace{0mm}
    \begin{lstlisting}[basicstyle=\ttfamily\scriptsize]
    1. void test(int x, int y) {
    2.    if (non_linear(y) == x) 
    3.      if (x > y + 10) ERROR; }
    \end{lstlisting}
    %\vspace{8.5mm}
    %\caption{}
  \end{subfigure}%
    \begin{subfigure}{.43\textwidth}
    %\vspace{-5.2mm}
    \begin{lstlisting}[basicstyle=\ttfamily\scriptsize]
      4. int non_linear(int v) {
      5.    return (v*v) % 50;
      6. }
    \end{lstlisting}
    %\vspace{3.5mm}
    %\caption{}
  \end{subfigure}%
  \end{center}
  \vspace{-4.0mm}
  \caption{Example with non-linear constraints.}
  \label{fi:non-linear-constraints}
  %\vspace{-2mm}
\end{figure}

\myparagraph{Concretization}
\cite{CS-CACM13} discusses limitations of classical symbolic execution in the presence of formulas that constraint solvers cannot solve, at least not efficiently. A concolic executor generates some random input for the program and executes it both concretely and symbolically: a possible value from the concrete execution can be used for a symbolic operand involved in a formula that is inherently hard for the solver, albeit at the cost of possibly sacrificing soundness in the exploration. 
%For instance, in the presence of three nested branches with only one being non-linear, {\sc DART}~\cite{DART-PLDI05} starts from a random valid input for the function, and then alters it when symbolically exploring the two linear branches. The work resorts to concretization also to avoid performing expensive or imprecise alias analysis on pointers. % with only one of them being

\boxedexample{In the code fragment of Figure~\ref{fi:non-linear-constraints}, the engine stores a non-linear constraint of the form $\alpha_x = (\alpha_y*\alpha_y)\,\%\,50$ for the $true$ branch at line 2. A solver that does not support non-linear arithmetic fails to generate any input for the program. However, a concolic engine can exploit concrete values to help the solver. For instance, if $x=3$ and $y=5$ are randomly chosen as initial input parameters, then the concrete execution does not take any of the two branches. Nonetheless, the engine can reuse the concrete value of $y$, simplifying the previous query as $\alpha_x = 25$ due to $\alpha_y = 5$. The straightforward solution to this query can now be used by the engine to explore both branches. Notice that if the value of $y$ is fixed to $5$, then there is no way of generating a new input that takes the first but not the second branch, inducing a false negative. In this case, a trivial solution could be to rerun the program choosing a different value for $y$ (e.g., if $y=2$ then $x=4$, which satisfies the first but not the second branch).   
}

% suggests to
To partially overcome the incompleteness due to concretization,~\cite{PRV-ISSTA11} suggests {\em mixed concrete-symbolic solving}, which considers {\em all} the path constraints collectable over a path before binding one or more symbols to specific concrete values. Indeed, {\sc DART}~\cite{DART-PLDI05} concretizes symbols based on the path constraints collected up to a target branch. In this manner, a constraint contained in a subsequent branch in the same path is not considered and it may be not satisfiable due to already concretized symbols. If this happen, {\sc DART} restarts the execution with different random concrete values, hoping to be able to satisfy the subsequent branch. The approach presented in~\cite{PRV-ISSTA11} requires instead to detect {\em solvable} constraints along a full path and to delay concretization as much as possible.

\myparagraph{Handling Problematic Constraints}
Strong SMT solvers allow executors to handle more path constraints directly, reducing the need to resort to concretization. This also results in a lower risk to incur a {\em blind commitment} to concrete values~\cite{DA-FSE14}, which happens when the under-approximation of path conditions from a random choice of concrete values for some variables results in an arbitrary restriction of the search space.
However, the decision problem for certain classes of constraints is well known to be undecidable, e.g., like for non-linear integer arithmetic, or the theory of reals with trigonometric functions often used to model real-world systems.
%\revedit{However, problems such as non-linear integer arithmetic or the theory of reals together with trigonometric functions are well known to be undecidable.} % SHORT VERSION
%Unfortunately, some constraints remain prohibitive for SMT solvers: for instance, non-linear integer arithmetic is undecidable in general; also, a branch condition might contain calls to opaque library methods such as trigonometric functions that would require special extensions to the solver to reason about.

\cite{DA-FSE14} proposes a {\em concolic walk} algorithm that can tackle control-flow dependencies involving non-linear arithmetic and library calls. The algorithm treats assignments of values to variables as a valuation space: the solutions of the linear constraints define a polytope that can be walked heuristically, while the remaining constraints are assigned with a fitness function measuring how close a valuation point is to matching the constraint. An adaptive search is performed on the polytope as points are picked on it and non-linear constraints evaluated on them. Compared to mixed concrete-symbolic solving~\cite{PRV-ISSTA11}, both techniques seek to avoid blind commitment. However, concolic walk does not rely on the solver for obtaining all the concrete inputs needed to evaluate complex constraints, and implements search heuristics that guide the walk on the polytope toward promising regions.

% Symcretic execution
% , which determines how close the branch conditions are to being satisfied and alters the concrete inputs to move closer to a full solution
%For instance, if an {\tt assert} statement is guarded by a branch condition that can be proven unsatisfiable, then there is no need to take into account all the other constraints along the path to the entry point to declare the target unreachable. A traditional concolic executor reasons instead about all the constraints along a path with a top-down approach, making it hard to detect the unreachability of a target statement because of constraints ``deep'' in the path.

\cite{DA-ASE14} describes {\em symcretic} execution, a novel combination of symbolic backward execution (SBE) (Section~\ref{se:executors}) and forward symbolic execution. The main idea is to divide exploration into two phases. In the first phase, SBE is performed from a target point and a trace is collected for each followed path. If any problematic constraints are met during the backward exploration, the engine marks them as {\em potentially} satisfiable by adding a special event to the trace and continues its reversed traversal. Whenever an entry point of the program is reached along any of the followed paths, the second phase starts. The engine concretely evaluates the collected trace, trying to satisfy any constraint marked as problematic during the first phase. This is done using a heuristic search, such as the concolic walk described above. An advantage of symcretic over classical concolic execution is that it can prevent the exploration of some unfeasible paths. For instance, the backward phase may determine that a statement is guarded by an unsatisfiable branch regardless of how the statement is reached, while a traditional concolic executor would detect the unfeasibility on a per-path basis only when the statement is reached, which is unfavorable for statements ``deep'' in a path.

%\myparagraph{Memory Page Size}
%In {\sc \stwoe}~\cite{CKC-TOCS12}, when a symbolic pointer is dereferenced, the engine determines which memory pages are referenced by it and passes their contents to the solver. As large page sizes can overwhelm the solver, {\sc \stwoe} uses small pages of configurable size rather than the default 4KB pages. The authors report significant performance benefits from using pages of smaller size.

%% file: hang.tex
% !TEX root = main.tex

\section{Further Directions}
\label{se:hang}

In this section we discuss how recent advances in related research areas could be applied or provide potential directions to enhance the state of the art of symbolic execution techniques. In particular, we discuss separation logic for data structures, techniques from the program verification and program analysis domains for dealing with path explosion, and symbolic computation for dealing with non-linear constraints.
% , i.e., polynomial constraints over variables

% Reynolds02,IO-POPL01
\subsection{Separation Logic}
%\mytempedit{
Checking memory safety properties for pointer programs is a major challenge in program verification. Recent years have witnessed {\em separation logic} (SL)~\cite{Reynolds02} emerging as one leading approach to reason about heap manipulations in imperative programs. SL extends Hoare logic to facilitate reasoning about programs that manipulate pointer data structures, and allows expressing complex invariants of heap configurations in a succinct manner.

At its core, a {\em separating conjunction} binary operator $*$ is used to assert that the heap can be partitioned in two components where its arguments separately hold. For instance, predicate $A * x\mapsto [n:y]$ says that there is a single heap cell $x$ pointing to a record that holds $y$ in its $n$ field, while $A$ holds for the rest of the heap.

Program state is modeled as a {\em symbolic heap} $\Pi\,\brokenvert\,\Sigma$: $\Pi$ is a finite set of pure predicates related to variables, while $\Sigma$ is a finite set of heap predicates. Symbolic heaps are SL formulas that are symbolically executed according to the program's code using an abstract semantics. SL rules are typically employed to support entailment of symbolic heaps, to infer which heap portions are not affected by a statement, and to ensure termination of symbolic execution via abstraction (e.g., using a widening operator).

A key to the success of SL lies in the local form of reasoning enabled by its $*$ operator, as it allows specifications that speak about the sole memory accessed by the code. This also fits together with the goal of deriving inductive definitions to describe mutable data structures. When compared to other verification approaches, the annotation burden on the user is rather little or often absent. For instance, the shape analysis presented in~\cite{CDO-JACM11} uses bi-abduction to automatically discover invariants on data structures and compute composable procedure summaries in SL.

% verification (Section~\ref{se:constraint-solving})
Several tools based on SL are available to date, for instance, for automatic memory bug discovery in user and system code, and verification of annotated programs against memory safety properties or design patterns. While some of them implement tailor-made decision procedures, \cite{BPS-ENTCS09,PWZ-CAV13} have shown that provers for decidable SL fragments can be integrated in an SMT solver, allowing for complete combinations with other theories relevant to program verification. This can pave the way to applications of SL in a broader setting: for instance, a symbolic executor could use it to reason inductively about code that manipulates structures such as lists and trees. While symbolic execution is at the core of SL, to the best of our knowledge there have not been uses of SL in symbolic executors to date.%We believe this might represent a promising research direction to follow.
%}

\subsection{Invariants} 
\label{ss:invariants}
Invariants are crucial for verifiers that can prove programs correct against their full functional specification. An invariant is a predicate true for an initial state and for each state reachable from it. Leveraging invariants can be beneficial to symbolic executors, in order to compactly capture the effects of a loop and reason about them. Unfortunately, we are not aware of symbolic executors taking advantage of this approach. One of the reasons might lie in the difficulty of computing loop invariants without requiring manual intervention from domain experts. In fact, lessons from the verification practice suggest that providing loop invariants is much harder compared to other specification elements such as method pre/post-conditions.

% a number of works [...] and might be
However, many researchers have recently explored techniques for inferring loop invariants automatically or with little human help~\cite{FMV-CSUR14}, which might be of interest for the symbolic execution community for a more efficient handling of loops. These approaches normally target inductive predicates, which are closed under the state transition relation (i.e., they make no reference to past behavior). Notice that all inductive predicates are invariants, but the converse is not true.

% that rank all -> over all
{\em Termination analysis} has been applied to verify program termination for industrial code: a formal argument is typically built by using one or more ranking functions over all the possible states in the program such that for every state transition, at least one function decreases~\cite{CPR-PLDI06}. Ranking functions can be constructed in a number of ways, e.g., by lazily building an invariant using counterexamples from traversed loop paths~\cite{GMR-PLDI15}. A termination argument can also be built by reasoning over transformed programs where loops are replaced with summaries based on transition invariants~\cite{TSW-TACAS11}. It has been observed that most loops in practice have relatively simple termination arguments~\cite{TSW-TACAS11}: the discovered invariants may thus not be rich enough for a verification setting~\cite{GFM-TSE15}. However, a constant or parametric bound on the number of iterations may still be computed from a ranking function and an invariant~\cite{GMR-PLDI15}.

{\em Predicate abstraction} is a form of abstract interpretation over a domain constructed using a given set of predicates, and has been used to infer universally quantified loop invariants~\cite{FQ-POPL02}, which are useful when manipulating arrays. Predicates can be heuristically collected from the code or supplied by the user: it would be interesting to explore a mutual reinforcing combination with symbolic execution, with additional useful predicates being originated during the symbolic exploration.

{\em LoopFrog}~\cite{LOOPFROG-ATVA08} replaces loops using a symbolic abstract transformer with respect to a set of abstract domains, obtaining a conservative abstraction of the original code. Abstract transformers are computed starting from the innermost loop, and the output is a loop-free summary of the program that can be handed to a model checker for verification. This approach can also be applied to non-recursive function calls, and might deserve some investigation in symbolic executors. 

Loop invariants can also be extracted using {\em interpolation}, a general technique that has already been applied in symbolic execution for different goals (Section~\ref{ss:interpolation}). %More generally, we believe that modern advances in invariant generation can provide potential solutions for handling loops more efficiently in a symbolic executor.
%
%}

% DROPPED as it's an application
\iffullver{
On the other hand, symbolic execution has proved useful to derive loop invariants. For instance, if a program contains an assertion after the loop, the approach presented in~\cite{PV-SPIN04} works backwards from the property to be checked and it iteratively applies approximation to derive loop invariants. The main idea is to pick the asserted property as the initial invariant candidate and then to exploit symbolic execution to check whether this property is inductive. If the invariant cannot be verified for some loop paths, it is replaced by a different invariant. The next candidate for the invariant is generated by exploiting the path constraints for the paths on which the verification has failed. Additional refinements steps are performed to guarantee termination. % [D] say weakness in not being able to find invariants that do not directly depends on the path conditions?
}{}

%Nevertheless, even symbolic execution can be used to derive loop invariants. Indeed, if a program contains an assertion after the loop, the approach presented in~\cite{PV-SPIN04} works backwards from the property to be checked and it iteratively applies approximation to derive loop invariants. The main idea is to pick the asserted property as the initial invariant candidate and then to exploit symbolic execution to check whether this property is inductive. If the invariant cannot be verified for some loop paths, it is replaced by a different invariant. The next candidate for the invariant is generated by exploiting the path constraints for the paths on which the verification has failed. Additional refinements steps are performed to guarantee termination.
%this can be exploited by a symbolic engine for automatically discovering some invariants over the loop. In~\cite{PV-SPIN04}, this is achieved by iteratively using \mynote{[D] Define?} invariant strengthening and approximation techniques. 

\subsection{Function Summaries}
%\mytempedit{
Function summaries (Section~\ref{ss:summarization}) have largely been employed in static and dynamic program analysis, especially in program verification. A number of such works could offer interesting opportunities to advance the state of the art in symbolic execution. For instance, the Calysto static checker~\cite{CALYSTO-ICSE08} walks the call graph of a program to construct a symbolic representation of the effects of each function, i.e., return values, writes to global variables, and memory locations accessed depending on its arguments. Each function is processed once, possibly inlining effects of small ones at their call sites. Static checkers such as Calysto and Saturn~\cite{SATURN-POPL05} trade scalability for soundness in summary construction, as they unroll loops only to a small number of iterations: their use in a symbolic execution setting may thus result in a loss of soundness. More fine-grained summaries are constructed in~\cite{RACERX-SOSP03} by taking into account different input conditions using a summary cache for memoizing the effects of a function.

\cite{SFS11} proposes a technique to extract function summaries for model checking where multiple specifications are typically checked one a time, so that summaries can be reused across verification runs. In particular, they are computed as over-approximations using interpolation (Section~\ref{ss:interpolation}) and refined across runs when too weak. The strength of this technique lies in the fact that an interpolant-based summary can capture all the possible execution traces through a function in a more compact way than the function itself. The technique has later been extended to deal with nested function calls in~\cite{SFS12}.%, which discusses an useful application in incremental update checking of programs.
%}

\subsection{Program Analysis and Optimization}
%\mytempedit{
We believe that the symbolic execution practice might further benefit from solutions that have been proposed for related problems in the programming languages realm. For instance, in the parallel computing community transformations such as {\em loop coalescing}~\cite{BGS-CSUR94} can restructure nested loops into a single loop by flattening the iteration space of their indexes. Such a transformation could potentially simplify a symbolic exploration, empowering search heuristics and state merging strategies. 

{\em Loop unfolding}~\cite{SK-SIGPLAN-NOTICES04} may possibly be interesting as well, as it allows exposing ``well-structured'' loops (e.g., showing invariant code, or having constants or affine functions as subscripts of array references) by peeling several iterations.

{\em Program synthesis} automatically constructs a program satisfying a high-level specification~\cite{PR-POPL89}. The technique has caught the attention of the verification community since~\cite{Solar-Lezama08} has shown how to find programs as a solution to SAT problems.
In Section~\ref{se:environment-thirdparty} we discussed its usage in~\cite{JQF-ICSE16} to produce compact models for complex Java frameworks: the technique takes as inputs classes, methods and types from a framework, along with tutorial programs (typically those provided by the vendor) that exercise its parts. We believe this approach deserves further investigation in the context of the path explosion problem. It could potentially be applied to software modules such as standard libraries to produce concise models that allow for a more scalable exploration of the search space, as synthesis can capture an external behavior while abstracting away entanglements of the implementation. %
%has shown that program synthesis can be used to build models for complex Java components by abstracting away the details and entanglements of their implementations while capturing their functional behavior. In particular, 

\subsection{Symbolic Computation}
%In particular, studies in the area of {\em symbolic computation}, also known as computer algebra,
Although the satisfiability problem is known to be NP-hard already for SAT, the mathematical developments over the past decades have produced several practically applicable methods to solve arithmetic formulas. In particular, advances in {\em symbolic computation} have produced powerful methods such as Gr\"{o}bner bases for solving systems of polynomial constraints, cylindrical algebraic decomposition for real algebraic geometry, and virtual substitution for polynomial real arithmetic formulas in which the degree of polynomials is no more than four~\cite{Abraham15}.

% handling complex Boolean constraints and; quantifier-free non-linear
% have proven to be
While SMT solvers are very efficient at combining theories and heuristics when processing complex expressions, they make use of symbolic computation techniques only to a little extent, and their support for non-linear real and integer arithmetic is still in its infancy~\cite{Abraham15}. To the best of our knowledge, only Z3~\cite{Z3-TACS08} and SMT-RAT~\cite{SMTRAT15} can reason about them both.
%provide support to reason about them both.

\cite{Abraham15} states that using symbolic computation techniques as theory plugins for SMT solvers is a promising symbiosis, as they provide powerful procedures for solving conjunctions of arithmetic constraints. The realization of this idea is hindered by the fact that available implementations of such procedures do not comply with the incremental, backtracking and explanation of inconsistencies properties expected of SMT-compliant theory solvers. One interesting project to look at is SC\textsuperscript{2}~\cite{SC2}, whose goal is to create a new community aiming at bridging the gap between symbolic computation and satisfiability checking, combining the strengths of both worlds in order to pursue problems currently beyond their individual reach.

Further opportunities to increase efficiency when tackling non-linear expressions might be found in the recent advances in {\em symbolic-numeric computation}~\cite{HandbookOfCompAlgebra}. In particular, these techniques aim at developing efficient polynomial solvers by combining numerical algorithms, which are very efficient in approximating local solutions but lack a global view, with the guarantees from symbolic computation techniques. This hybrid techniques can extend the domain of efficiently solvable problems, and thus be of interest for non-linear constraints from symbolic execution.

%% file: conclusions.tex
% !TEX root = main.tex

\section{Conclusions}
\label{se:conclusions}

Symbolic execution techniques have evolved significantly in the last decade, with notable applications to compelling problems from several domains like software testing (e.g., test input generation, regression testing), security (e.g., exploit generation, authentication bypass), and code analysis (e.g., program deobfuscation, dynamic software updating). This trend has not only improved existing solutions, but also led to novel ideas and, in some cases, to major practical breakthroughs. For instance, the push for scalable automated program analyses in security has culminated in the 2016 DARPA Cyber Grand Challenge, which hosted systems for detecting and fixing vulnerabilities in unknown software with no human intervention, such as {\sc Angr}~\cite{ANGR-SSP16} and {\sc Mayhem}~\cite{MAYHEM-SP12}, that competed for nearly \$4M in prize money.

%\noindent
This survey has discussed some of the key aspects and challenges of symbolic execution, presenting for a broad audience the basic design principles of symbolic executors and the main optimization techniques. We hope it will help non-experts grasp the key inventions in this exciting line of research, inspiring further work and new ideas.

%\ifx\arxivver\undefined
\specialcomment{online}{
\begingroup
\subsection*{ELECTRONIC APPENDIX}
\phantomsection\addcontentsline{toc}{subsection}{Electronic Appendix}
}{%
\endgroup
}
%\fi

%\begin{online}
\subsection*{ELECTRONIC APPENDIX}
The online appendix of this manuscript discusses a selection of prominent applications of symbolic execution techniques, addresses further challenges that arise in the analysis of programs in binary form, and provides a list of popular symbolic engines.
%\end{online}

\ifdefined\arxivver
\subsection*{ACKNOWLEDGEMENTS}
We thank the anonymous referees for their valuable comments and helpful suggestions. This work is supported in part by a grant of the Italian Presidency of the Council of Ministers and by the CINI (Consorzio Interuniversitario Nazionale Informatica) National Laboratory of Cyber Security. % (Consorzio Interuniversitario Nazionale Informatica) 
\else
\begin{acks}
We thank the anonymous referees for their valuable comments and helpful suggestions. This work is supported in part by a grant of the Italian Presidency of the Council of Ministers and by the CINI (Consorzio Interuniversitario Nazionale Informatica) National Laboratory of Cyber Security. % (Consorzio Interuniversitario Nazionale Informatica) 
\end{acks}
\fi

\iffalse
Techniques for symbolic execution have evolved significantly in the last decade, leading to major practical breakthroughs. In 2016, the DARPA Cyber Grand Challenge hosted systems that can detect and fix vulnerabilities in unknown software with no human intervention, such as {\sc Angr}~\cite{ANGR-SSP16} and {\sc Mayhem}~\cite{MAYHEM-SP12}, which won the \$2M first prize. {\sc Mayhem} was also the first autonomous software to play the Capture-The-Flag contest at the DEF CON 24 hacker convention\footnote{\url{https://www.defcon.org/html/defcon-24/dc-24-ctf.html}.}. The event demonstrated that tools for automatic exploit detection based on symbolic execution can be competitive with human experts, paving the road to unprecedented applications %and the rise of start-ups 
that have the potential to shape software %security and 
reliability in the next decades. 

This survey has discussed some of the key aspects and challenges of symbolic execution, presenting them for a broad audience. 
To explain the basic design principles of symbolic executors and the main optimization techniques, we have focused on single-threaded applications with integer arithmetic. Symbolic execution of multi-threaded programs is treated, e.g., \iffullver{in~\cite{KPV-TACAS03,SA-HVC06,CLOUD9-EUROSYS11,FHR-ESEC13,BGC-OOPSLA14,GKW-ESEC15}}
{in~\cite{BGC-OOPSLA14,GKW-ESEC15}}, 
%{in~\cite{FHR-ESEC13,BGC-OOPSLA14,GKW-ESEC15}}, 
while techniques for programs that manipulate floating point data are addressed \iffullver{in, e.g., \cite{M-STVR01,BGM-STVR06,LTH-ICTSS10,CCK-EUROSYS11,BVL-POPL13,CCK-TSE14,RPW-SIGSOFT15}}
{in, e.g., \cite{RPW-SIGSOFT15}}.
%{in, e.g., \cite{BVL-POPL13,CCK-TSE14,RPW-SIGSOFT15}}.

We hope that this survey will help non-experts grasp the key inventions in the exciting line of research of symbolic execution, inspiring further work and new ideas.
\fi

%\myparagraph{Acknowledgements}
%This work is partially supported by a grant of the Italian Presidency of Ministry Council and by the CINI  (Consorzio Interuniversitario Nazionale Informatica) Cybersecurity National Laboratory.
%This work is supported in part by a grant of the Italian Presidency of the Council of Ministers and by the CINI (Consorzio Interuniversitario Nazionale Informatica) National Laboratory of Cyber Security.

\ifdefined\arxivver
\myparagraph{Live Version of this Article}
We complement the traditional scholarly publication model by maintaining a live version of this article at {\href{https://github.com/season-lab/survey-symbolic-execution}{https://github.com/season-lab/survey-symbolic-execution/}}. The live version incorporates continuous feedback by the community, providing post-publication fixes, improvements, and extensions.
\fi